\documentclass{cs19proc}

\editors{G.~A. Feiden}
\publisher{Zenodo}
\conference{The 19th Cambridge Workshop on Cool Stars, Stellar Systems, and the Sun}
\conferencedate{2016}

\title{Mass Losing Asymptotic Giant Branch Stars and Supergiants}
\author{Patricia A. Whitelock$^{1,2}$, Martha Boyer$^3$, Susanne H\"ofner$^4$, Markus Wittkowski$^5$ and Albert A. Zijlstra$^6$ }

\affiliation{$^{1}$ South African Astronomical Observatory, P O Box 9, 7935 Observatory, South Africa \\
$^{2}$ Department of Astronomy, University of Cape Town, 7701 Rondebosch, South Africa \\
$^3$ STScI, 3700 San Martin Drive, Baltimore, MD 21218, USA\\
$^{4}$ Department of Physics and Astronomy, Uppsala University, Box 516, SE-75120 Uppsala, Sweden\\
$^5$ ESO, Karl-Schwarzschild-Str. 2, 85748 Garching, Germany \\
$^6$ Jodrell Bank Centre for Astrophysics, School of Physics \& Astronomy, University of Manchester, Oxford Road, Manchester, M13 9PL, UK}

\shorttitle{AGB stars and Supergiants}
\shortauthors{Whitelock et al.}

\abs{This paper presents a summary of four invited and twelve contributed presentations on asymptotic giant branch stars and red supergiants, given over the course of two afternoon splinter sessions at the 19th Cool Stars Workshop. It highlights both recent observations and recent theory, with some emphasis on high spatial resolution, over a wide range of wavelengths.    Topics covered include 3D models, convection, binary interactions, mass loss, dust formation and magnetic fields.}

\begin{document}

\maketitle

\section{Introduction}
There have been extraordinary advances in our knowledge of asymptotic giant branch (AGB) stars over the last decade. On the observational side Spitzer, Herschel and ALMA in particular have provided access to the wavelength ranges in which these stars and their associated dust and molecular shells emit most of their energy. Interferometry has enabled convection cells to be resolved and has highlighted the role of binary interactions in the mass-loss process from these huge stars. At the same time theoretical advances give us a better understanding of element formation, 3D models of convection, and new insight into the properties of grains produced in the very extended circumstellar environments (see invited presentation by H\"ofner to the main meeting). Nucleosynthesis models are making testable
predictions and population synthesis models are reproducing many of the characteristics of highly evolved stars, for the first time (see invited presentation by Karakas to the main meeting).

In this two day splinter session we covered some of the recent observational and theoretical advances in the understanding of AGB stars and red supergiants (RSG), as well as touching on many aspects that remain puzzling. We nominally divided the two days so that on the first we focused on the star itself, and discuss pulsation, convection, surface magnetic fields etc.; while on the second we examined the circumstellar environment, dust formation, binary interactions etc. In practice the star and its environment are so closely linked that there is considerable overlap in the topics discussed in the two sessions.

\section{Day 1: From waves to winds: gas dynamics in evolved stars}
Luminous cool giants are strongly affected by dynamical processes. Large-scale convective motions and stellar pulsation trigger sound waves, which may develop into strong shock waves as they propagate outwards through the stellar atmospheres. These shocks, possibly supported by other physical processes (rotation, magnetic fields), intermittently lift gas to distances where dust grains can form and trigger a stellar wind. Recent developments in instrumentation allow us to image the relevant regions in ever increasing detail over a wide wavelength range, from the optical to the radio regime (e.g., VLTI, CHARA, ALMA), giving unprecedented insights into these phenomena. At the same time, dynamical models of interior dynamics (3D convection, pulsation) and of atmospheres and winds have reached a level where they permit self-consistent quantitative simulations of these processes. Significant progress can be expected from a detailed comparison of the latest 3D models and imaging observation.

\begin{figure*}
	\centering
	\resizebox{\hsize}{!}{
	\includegraphics[width=0.29\linewidth]{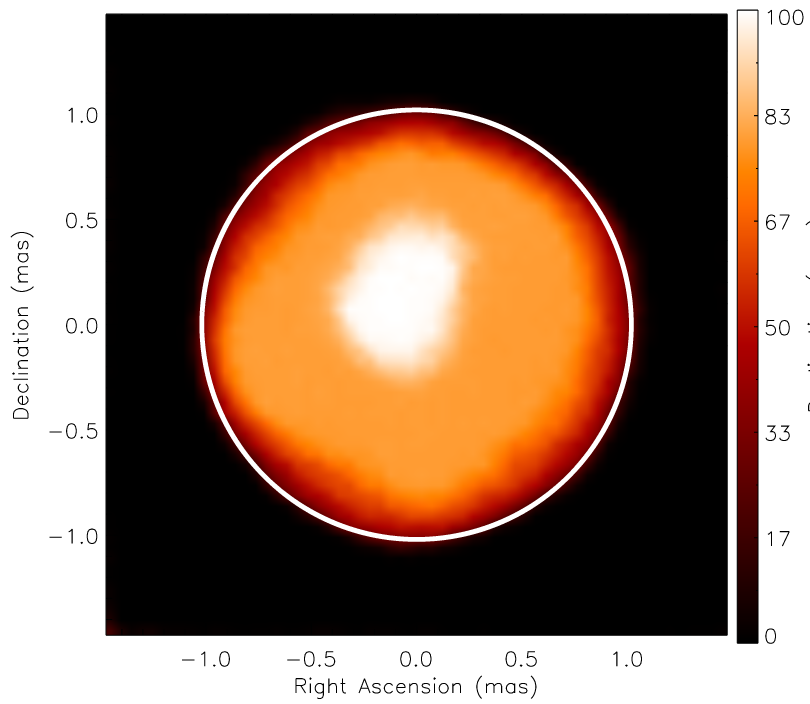}
	\includegraphics[width=0.26\linewidth]{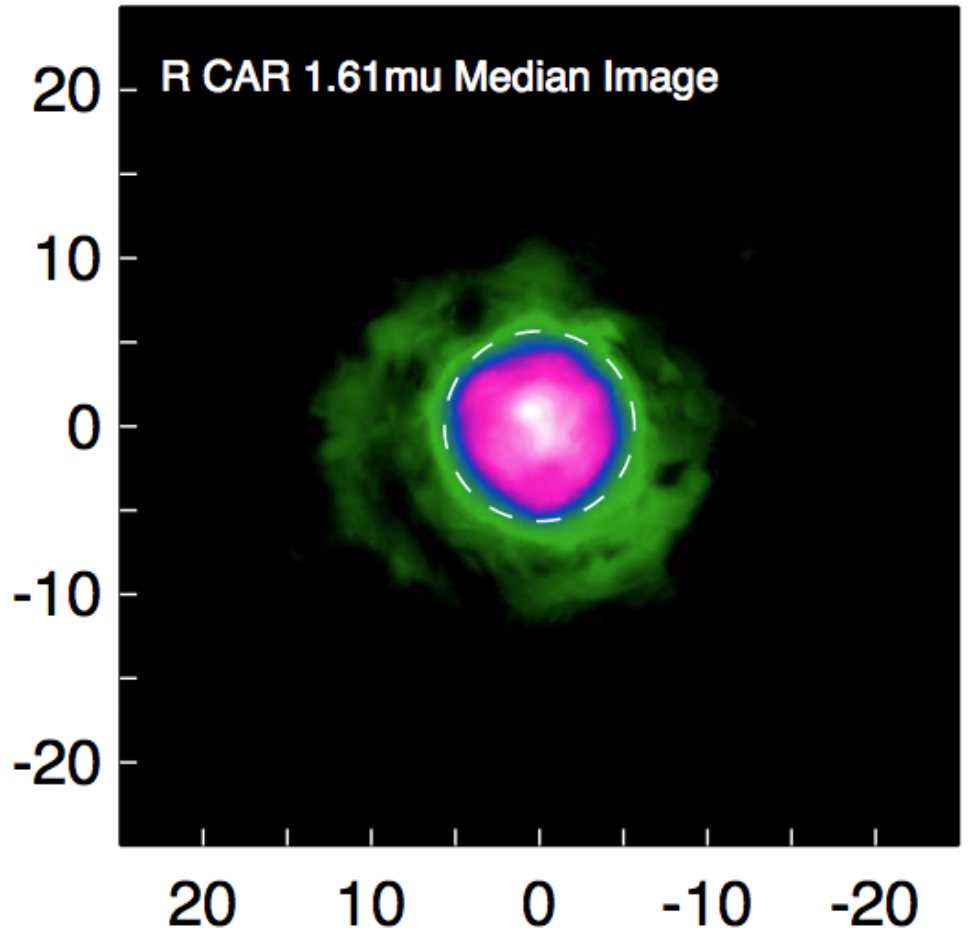}
	}
	\caption{Left Panel: interferometric image of the RSG star T~Per obtained with the CHARA array \citep[courtesy of F. Baron,][]{baron2014}. Right Panel: interferometric image of the
	oxygen-rich Mira variable R~Car obtained with the VLTI array  \citep[courtesy of J.~Monnier,][]{monnier2014}. 
For both panels the axes units are milli-arcsec.}
	\label{image}
\end{figure*}

\subsection{The coolest surfaces we have seen so far in the infrared}

Claudia Paladini\footnote{Institut d'Astronomie et d'Astrophysique, Brussels, Belgium} started the opening invited presentation by pointing out that although the mass-loss process from evolved stars has been under investigation since the early 1970s there are still many uncertainties associated with it. Open questions include: (i) what drives the winds for RSG stars, and for O-rich and small-amplitude pulsating C-rich AGB stars, (ii) how does the stellar wind affect the various spatial scales of the stellar atmosphere, from the photosphere to the interface with the stellar envelope. She briefly touches on the first question, and concentrates on the second one, with a focus on what has been learned by studying the inner spatial scales (1-10~$R_\star$) that are probed with long-baseline interferometry. 

%%if you want you can remove this part%%
The first interferometric studies involved the combination of only two or three telescopes.  Detecting asymmetric structures was quite tricky at that time and most of the earlier studies concentrate on stellar parameter determination, variability studies, and on providing constraints for stellar atmosphere models \citep[see section \ref{hron} and][for a recent example]{Wittkowski2016}. Among interesting recent results on stratification, \citet{arroyo-torres2013, arroyo-torres2014, Arroyo2015} showed how 
PHOENIX hydrostatic models are too compact to reproduce the observed extension of RSG stars in the near-infrared.
%while on the other hand they can reproduce the observations of small amplitude AGB stars. 
The use of 3D models including a convection prescription does not improve the situation for RSGs, and even the 1D CODEX models implementing pulsation are not able to levitate the molecular layers to the observed extension. Magnetic fields have been measured on the surface of a RSG \citep{auriere2010}. Ultimately, perhaps all these mechanisms have to be taken into account for effective modelling of RSG stars. Within this framework, simultaneous interferometric and polarimetric observations will help to clarify the interplay between magnetic field and convection \citep{Montarges2016, Auriere2016}.
%%until here%%

Despite the challenging nature of the observations, already in the late 1990s and early 2000, various high angular resolution data pointed to departures from spherical symmetry \citep{wittkowski2014}. For many years the interpretation of such structures was not unequivocal and involved (i) spots due to convection, or (ii) elliptical distortion due to stellar rotation \citep{vanbelle2013, cruzalebes2015}. However, at least in the case of AGB stars, high stellar rotation rates are usually connected to the presence of a binary companion transferring momentum to the primary \citep{barnbaum1995, mayer2014}.

\begin{figure*}
	\centering
	\includegraphics[width=0.85\linewidth]{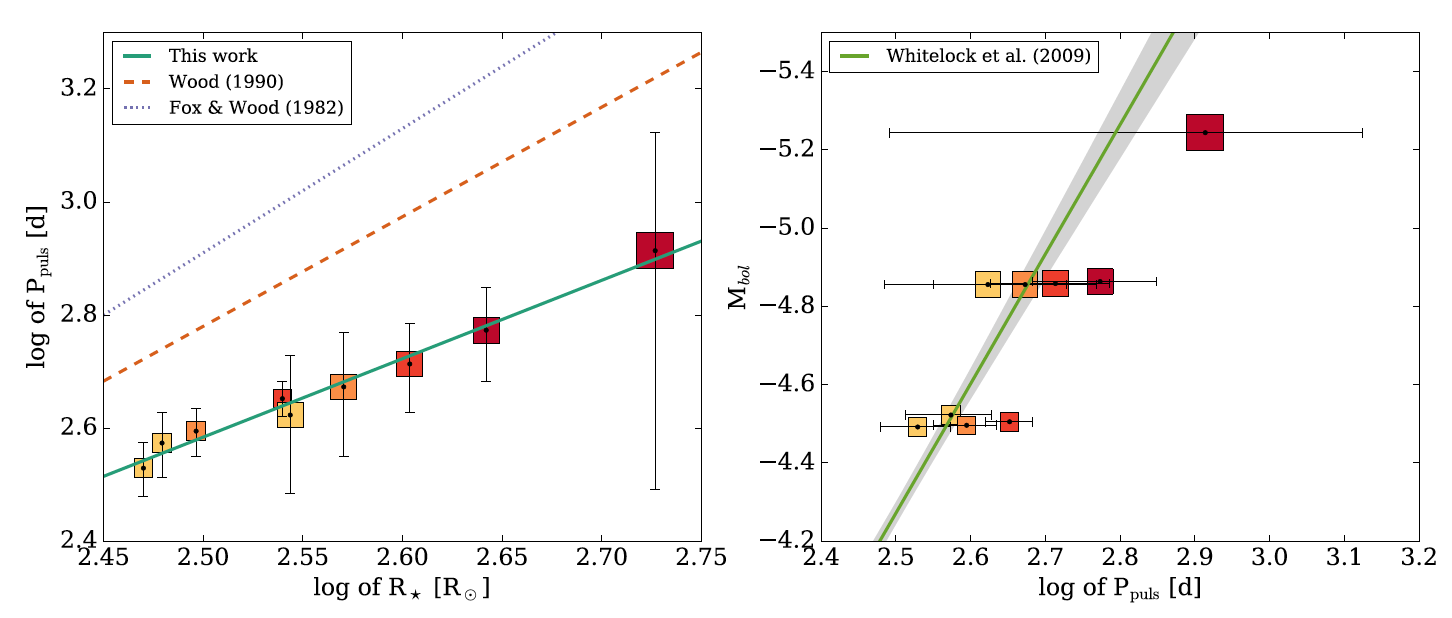}
	\caption{A comparison of 3D and 1D pulsation models (left) and of 3D models with observations (right). The colours of the symbols indicate the effective temperature of the model (red-cooler, yellow-hotter), while the sizes of the symbols represent the luminosity (large-10 000, medium-7000, small-5000 $L_\odot$) (Freytag, Liljegren \& H\"ofner  A\&A submitted). }
	\label{fig:liljegren}
\end{figure*}

\citet{schwarzschild1975} predicted, and more recently \citet{freytag2008} and \citet{chiavassa2015} demonstrated, how the number of convective cells on the stellar surface decreases as the star reaches late evolutionary stages. 
While the Sun shows up to $10^7$ granules, the number of convective cells drops to only a few for AGB and RSG stars. Moreover, AGB surfaces are expected to be more irregular than those of RSGs. Convection appears to be the most likely explanation for the measured deviation from symmetry, but we had to wait for interferometry to deliver the first images to be sure. The first interferometric images of Betelgeuse \citep{Haubois2009, chiavassa2010b}, and more recently images of other RSGs presented by \citet{chiavassa2010a}, \citet{baron2014} and \citet{monnier2014} show indeed at least one prominent surface structure in each star (left panel of Fig.~\ref{image}). 

The surfaces of RSGs appear mostly symmetric in the near-infrared, 
but one should keep in mind that, because of the lack of complete information in the uv-coverage, some of these ``first generation'' interferometric images are model dependent.
Interferometric images of AGB stars have been presented by \citet{ragland2008}, \citet{lacour2009}, \citet{lebouquin2009} and \citet{haubois2015}. These works show the presence of molecular shells, sometimes asymmetric, but no obvious convective patterns were detected. The main reason was again poor uv-coverage because of the use of 3-beam combiner interferometers. 
As a consequence, strong model dependency is involved in all of these interferometric image reconstructions.  

In January 2014 data for the O-rich Mira R~Car were recorded during technical time with the $H$-band 4-telescope beam combiner VLTI/PIONIER.
The data were used for the 2014 image reconstruction contest \citep{monnier2014}. The surface of R~Car shows a bright spot, and the atmosphere appears less homogeneous than in the
RSG case (right panel, Fig.~\ref{image}). At the time of the CS19 meeting, at least 3 more AGB stars have been imaged with PIONIER. Preliminary results qualitatively confirm the model predictions, and suggest that the atmospheres of carbon AGB stars are more complex than those of their oxygen-rich counterparts (Paladini {\it et al.}, prep.; Wittkowski {\it et al.}, prep.). Starting from September 2016 VLTI/GRAVITY will deliver images of AGB and RSG stars across the $K$-band, with 4000 spectral resolution elements. This will allow us to map the continuum as well as the molecular layers, and possibly to characterize the convective patterns in more  detail.

Observations in the mid-infrared enable us to move further from the stellar surface, and to probe the dust-forming region. The interferometers available until very recently involved (again) combining the beams from only two or three telescopes. To the best of Paladini's knowledge no image reconstruction has been attempted using long-baseline interferometric data. Departure from spherically symmetric geometry has been observed in the atmospheres of both AGB and RSG stars; however, it is not yet clear if the asymmetric structures are the signature of randomly distributed dust clumps \citep{chandler2007, ravi2011, paladini2012,sacuto2013}, or circumstellar/circum-binary discs \citep{deroo2007, ohnaka2008a, klotz2012}. Single-dish observations show that both discs and dust-plumes are present around these stars at spatial scales larger than 5-10~R$_\star$, but only images with the second generation VLTI/MATISSE interferometer will help clarify what is happening deeper inside the dust forming region.

In conclusion, convection plays a crucial role on the surfaces of AGB and RSG stars, but magnetic fields might also play a role in shaping the environment of these stars
between 1-3~$R_\star$. In the dust-forming region dust plumes are detected, and imaging studies with the new generation of interferometers might reveal
previously unseen binaries.
From the observational point of view, one will need (i) monitoring studies to be able to quantify the lifetime of the convection patterns on the stellar surface, (ii) coordinated studies using different techniques to be able to understand the interplay between convection and magnetic field. The latter question might be most easily tackled with an interferometer with a polarimetry mode.
We are at the stage where images of stellar surfaces are becoming more  routine, and therefore from the theoretical point of view grids of 3D models are required for a detailed comparison.
From 2019 VLTI/MATISSE will deliver images at 2-10 $R_\star$ in the $L$, $M$, and $N$ bands, i.e., right in the middle of the dust-forming region. At that stage, the community would benefit from simultaneous observations with GRAVITY and MATISSE in order to study efficiently both the stratification and the variability of the various structures.

\subsection{Pulsation Properties of an AGB star 3-D model grid }
Sofie Liljegren\footnote{Uppsala Universitet, Sweden} presented results on behalf of herself, Bernd Freytag and Susanne H\"ofner. 
She first noted that AGB stars are evolved stars with low to intermediate
mass and although they are thought to be important for galactic chemical evolution,
current theoretical models are not complete and do not explain all available
observations. To improve on this situation, the first ever grid of 3D star-in-a-box simulations of
AGB stars has been calculated (Freytag, Liljegren \& H\"ofner, A\&A submitted) and is briefly described below.

The CO5BOLD radiation hydrodynamical code was used to simulate the interior and
the inner atmosphere of the AGB stars. The model mass was set to one solar mass and the influence of other stellar parameters explored via a grid that contains eight models with effective temperatures ranging
from 2500K to 2800K and luminosities from 5000 to 10000 solar luminosities.  With several models of different stellar
parameters it is possible to extract pulsation properties, such as period and amplitude,
and investigate the resulting trends.

The results are illustrated in Fig.~\ref{fig:liljegren}. This shows that
the 3D models generally give a larger radius for a given period (left in Fig.\ref{fig:liljegren}), when
compared to 1D pulsation models (see, e.g., \citet{Wood1990,FoxWood1982}).
%e.g. Wood, 1990, Fox \& Wood 1982). 
When compared to an observed period-luminosity relationship (right)  \citep{Whitelock2009}  there is a good fit.
These results indicate that the 3D models give a satisfactory description of the
stellar interior, and could be used to investigate the interplay between self-excited
pulsation, shocks and dust formation.

\begin{figure*}
\includegraphics[width=0.46\textwidth]{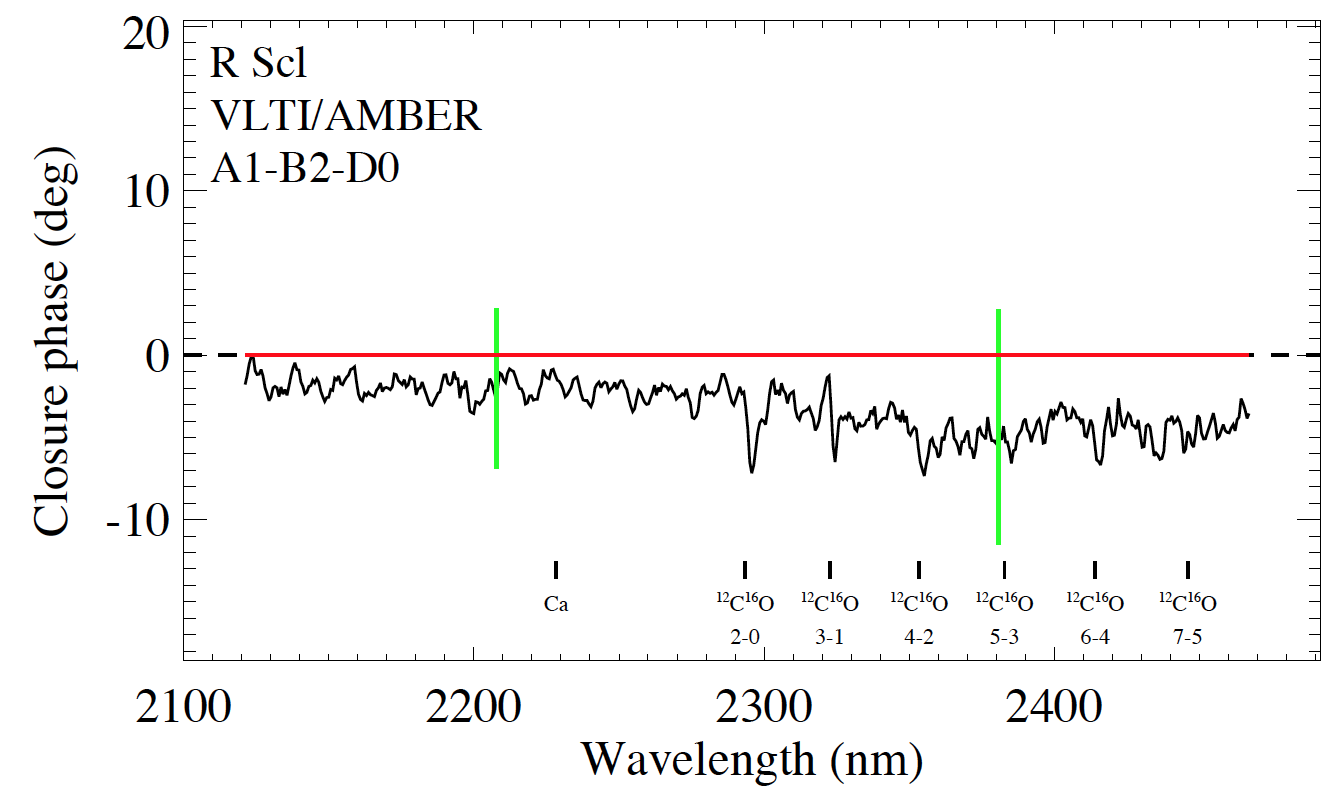}
  \includegraphics[width=0.46\textwidth]{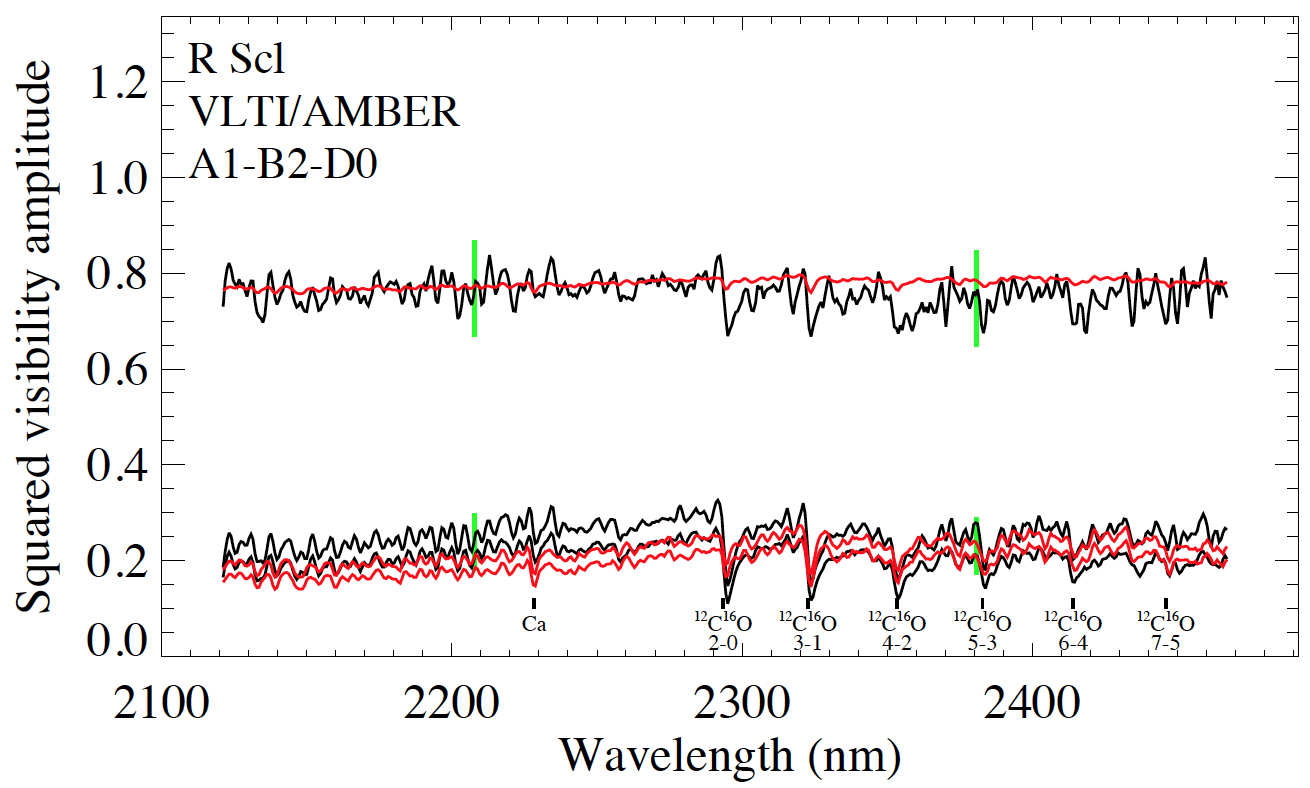}
\caption{Preliminary AMBER visibility results as a function 
of wavelength (black)
compared to the prediction from the best-fit model atmosphere (red)
for one AMBER data set taken on 2012-10-17
with the compact VLTI baseline configuration A1/B2/D0.
The green bars denote the
mean errors for the first and second halves of the wavelength range.}
\label{fig:amber_visspectra}
\end{figure*}

\subsection{VLTI/PIONIER imaging observations of the AGB star R Sculptoris}
Markus Wittkowski\footnote{ESO Garching, Germany} discussed R~Scl, a carbon-rich semi-regular AGB star
with a pulsation period of 370 days \citep{Samus2009}.
\citet{Maercker2012} obtained ALMA images of the CO (3-2 and 2-1) emission
from R~Scl. These images surprisingly revealed an inner spiral structure
 connected to a known detached CO shell \citep{Olofsson1988,Olofsson2010},
indicating a previously undetected binary companion.
\citet{Maercker2012} explained these
structures as the consequence of a variable mass-loss history following a thermal pulse
1800 years ago, together with a binary companion. Their model has an orbital period of
350 years and a separation of 60 AU, assuming a distance of 290~pc, and
employs a hydrodynamic simulation.

Wittkowski presented preliminary results from $K$-band 
interferometry  of R~Scl obtained with the VLTI instrument AMBER and 
$H$-band interferometric imaging obtained with the
instrument PIONIER. The goals of this study include (1) further constraining
and testing the available dynamic atmosphere and wind models with
near-infrared interferometry, (2) revealing the detailed morphology
of the stellar atmosphere and innermost mass-loss region
and (3) constraining fundamental stellar
properties of R~Scl.

Figure~\ref{fig:amber_visspectra} shows the squared visibility
amplitudes and closure phases as a function of wavelength.
Synthetic values based on the best-fit model atmosphere 
\citep{Mattsson2010,Eriksson2014} are also shown. 
The observed squared visibility
values show drops at the positions of the $^{12}$C$^{16}$O bandheads.
This is best seen at squared visibility levels of 0.1--0.2
and indicates that these lines are formed in extended layers above
the photosphere. This behaviour is comparable to that seen in oxygen-rich
semi-regular AGB stars \citep[e.g.][]{Marti-Vidal2011} or
Miras \citep[e.g.][]{Wittkowski2016}. The observed visibility values
are consistent with the model prediction.
Closure phases are consistent
with zero, given the relatively large errors. However, the errors
are dominated by systematic effects, and the wavelength-differential
errors are much smaller. The latter show closure phase features
at the positions of the CO bandheads, indicating photocentre
displacements between the CO-forming regions and the nearby
pseudo-continuum.

\begin{figure}
\center
\includegraphics[width=4.5cm]{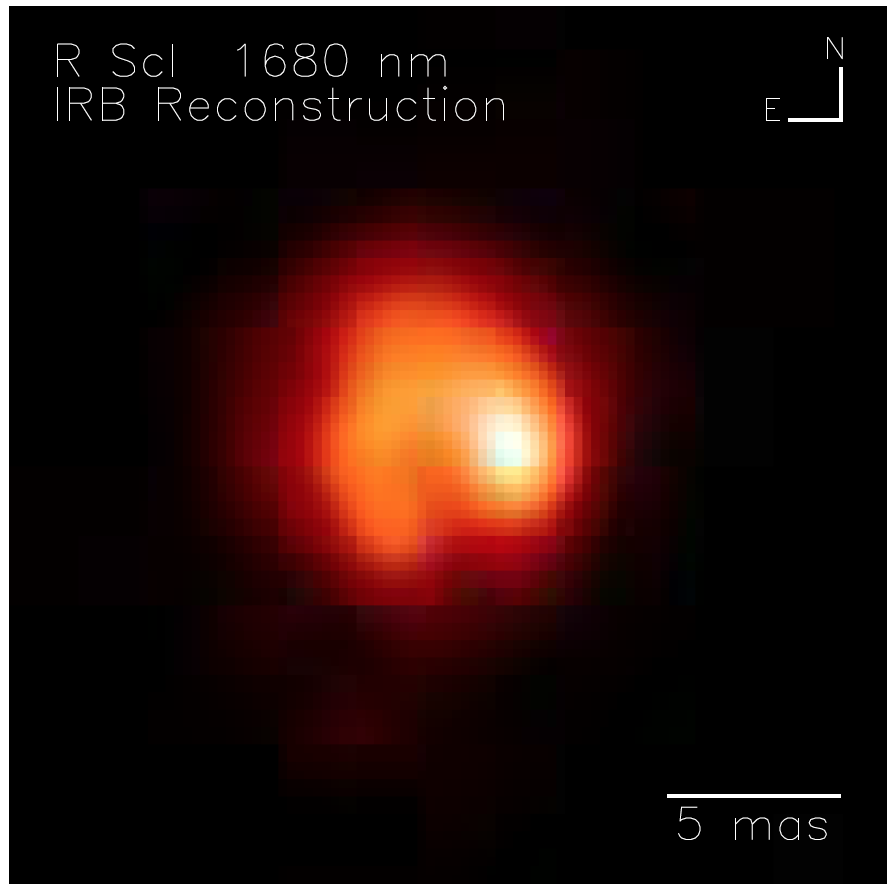}
\caption{Preliminary image reconstruction of the surface structure
of R~Scl based on data obtained with the VLTI/PIONIER
instrument in 2014 in the $H$-band,  reconstructed
with the IRBis package of \protect\citet{Hofmann2014}.
For comparison, the angular Rosseland photospheric diameter 
is estimated to be about 9\,mas (the bar in the lower right corner is 5\,mas). }
\label{fig:image}
\end{figure}

Figure ~\ref{fig:image} shows a preliminary image reconstruction
of the surface of R~Scl in the $H$-band. 
In this image, R Scl shows one dominant surface spot and a spiral-like
structure. Most likely these features are a consequence 
of convection. Less likely, the binary 
companion may produce effects very close to the stellar surface.
For comparison the effects of Mira~B on the circumstellar environment
of Mira~A have been observed at a distance
of about 10 stellar radii or less \citep{Ramstedt2014,Wong2016}.
The details of these VLTI observations will be described in a
paper by Wittkowski {\it et al.} (in preparation).

\subsection{Wind models for M-type AGB stars in the Large Magellanic Cloud}
Sara Bladh\footnote{University of Padova, Italy} noted that the
stellar winds observed in evolved AGB stars are usually attributed to a combination of stellar pulsations and radiation pressure on dust. Shock waves triggered by pulsations propagate through the atmosphere, compressing the gas and lifting it to cooler regions, thereby creating favorable conditions for grain growth. If sufficient radiative acceleration is exerted on the newly formed grains, either through absorption or scattering of stellar photons, an outflow can be triggered. 

Time-dependent wind models (DARWIN models, see \citet{Hofner2016}
%Hoefner et al. 2016), 
using this scenario for the mass loss have successfully been able to produce stellar outflows with dynamical properties compatible with observations for both C-type and M-type AGB stars \citep[e.g.][]{Eriksson2014,Bladh2015}.
%(e.g. Eriksson et al. 2014, Bladh et al. 2015).
 In the wind models for C-type AGB stars the outflows are predominately driven by photon absorption on amorphous carbon grain, while in the models for M-type AGB stars the stellar winds are triggered by photon scattering on micron-sized Mg-rich silicates.

\begin{figure}
\centering
\includegraphics[width=8cm]{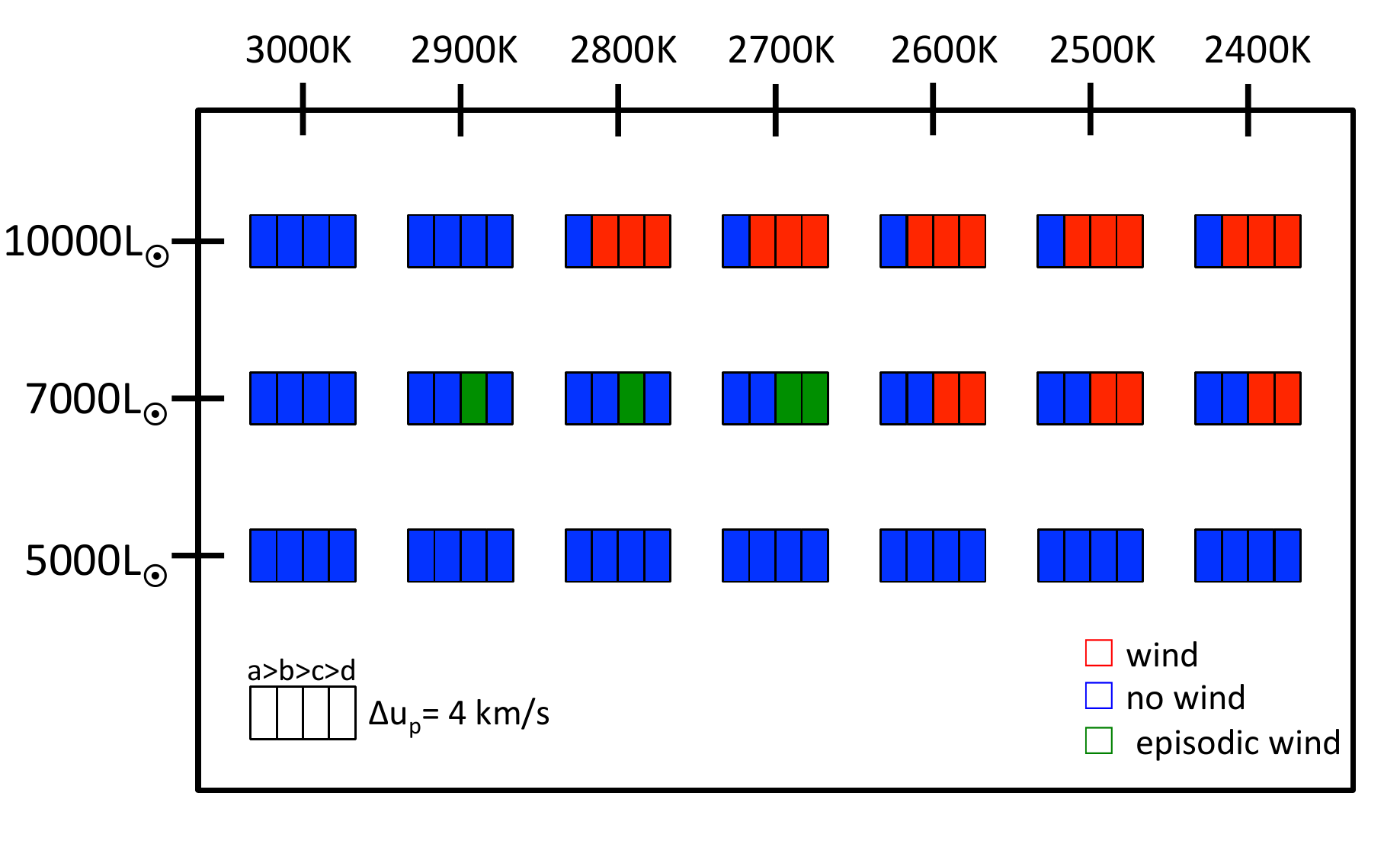}
   \caption{Schematic overview of the dynamical properties of the DARWIN models with LMC metallicity as a function of input parameters. The red boxes represent models that develop a stellar wind, the blue boxes represent models where no wind forms and the green boxes indicate models with periodic mass loss. The pulsation period of these models is derived from the period-luminosity relation in \citet{Feast1989} and the piston velocity amplitude is set to $\rm 4\,km\,s^{-1}$. The seed particle abundance $n_d/n_H$ is increasing towards the right, $[a,b,c,d]=[1\times10^{-15},3\times10^{-15},1\times10^{-14},3\times10^{-14}]$.}
    \label{f_dyn1}
\end{figure}

What happens to the wind properties of AGB stars in a low metallicity environment such as the Large Magellanic Cloud (LMC) or Small Magellanic Cloud (SMC)? C-type AGB stars produce their own carbon during the AGB phase and the mass-loss rates should therefore not be significantly affected by metallicity. The outflows in M-type AGB stars, however, are driven by dust material consisting of elements that cannot be produced by the stars themselves. A low metallicity environment could for that reason have a strong impact on the mass loss of these stars.
 
A recent investigation by Bladh and colleagues of the properties of M-type AGB stars in a low metallicity environment has provided the first tentative results from a set of DARWIN models for M-type AGB stars with metallicity similar to that of the LMC. The grid is set up much like the previous grid of wind models with solar metallicity \citep{Bladh2015}. The models in this set have solar mass, three different stellar luminosities (5000\,L$_{\odot}$, 7000\,L$_{\odot}$, and 10000\,L$_{\odot}$) and an effective temperature that ranges between 2400\,K and 3000\,K. A schematic overview of the dynamical properties of this set is shown in Fig.~\ref{f_dyn1}. 

\begin{figure}
\centering
\includegraphics[width=8cm]{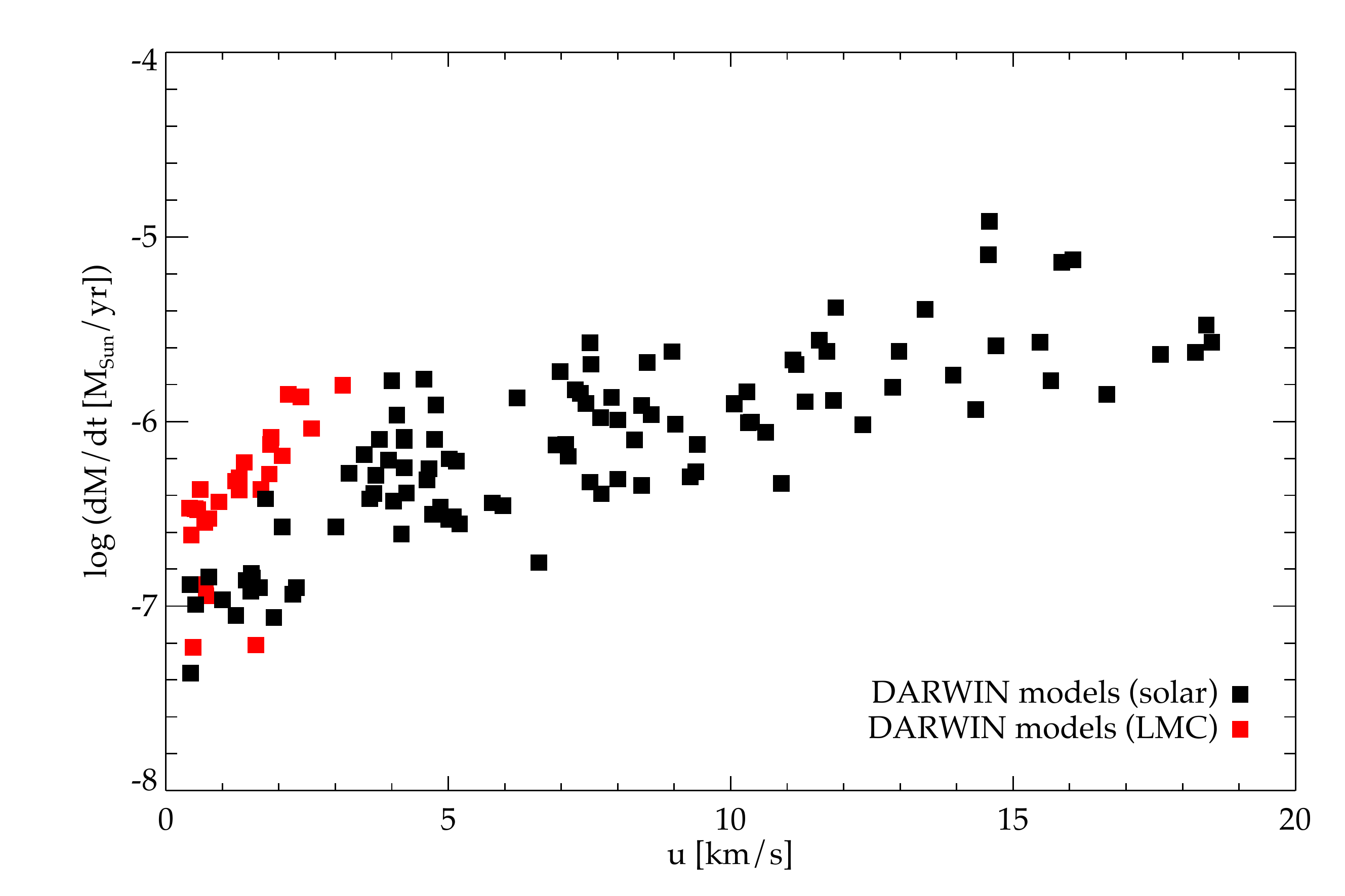}
   \caption{Mass-loss rates vs. wind velocities for DARWIN models with LMC metallicity (red) and solar metallicity (black).}
    \label{f_dyn2}
\end{figure}

The dynamical properties of this set verify the assumption that a low metallicity environment affects the mass loss of M-type AGB stars: the models produce stellar winds only for very luminous AGB stars and, as can be seen in Fig.~\ref{f_dyn2}, the resulting wind velocities are very low ($\rm < 5\,km\,s^{-1}$), but the total, gas plus dust, mass-loss rates are comparable to those produced from the corresponding grid for solar metallicity.

\subsection{AGB and Red Supergiant Stars: From the Radio to the Submillimetre}

\begin{figure*}
\centering
\includegraphics{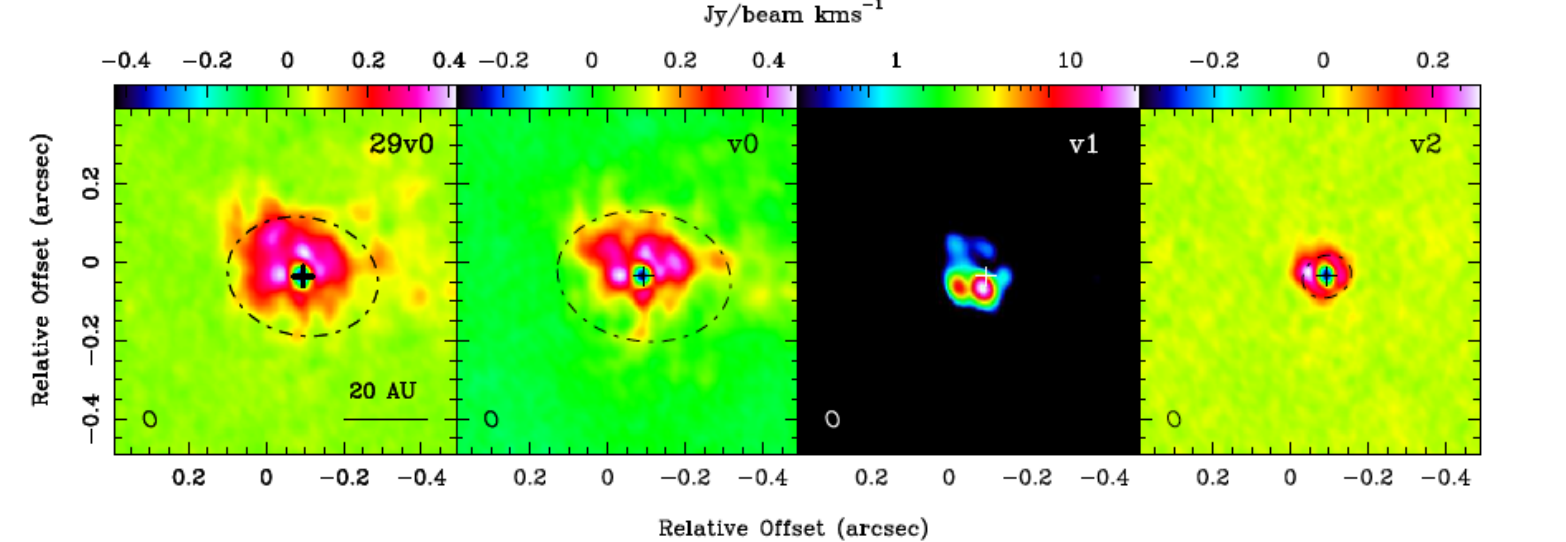}
   \caption{ALMA long baseline observations of different SiO J=5-4 lines towards Mira (Wittkowski, Humphreys {\it et al.} in preparation). }
    \label{fig:humphries}
\end{figure*}

Liz Humphreys\footnote{ESO Garching, Germany} started her invited review by noting that radio and submillimetre interferometry have enabled the study of AGB and RSG stars from the stellar photosphere to the outer wind, at up to sub-milli-arcsecond resolution. She focused on VLBI and ALMA observations of the gas dynamics of the stellar extended atmospheres/inner circumstellar envelopes (CSEs), where shocks are believed to play an important role in levitating gas to larger radii, enabling dust formation. She also discussed submm continuum and maser observations that provide indications of asymmetry, clumpiness and inhomogeneity in the inner CSE, with mass loss occurring in localised directions. Finally, she presented evidence for dynamically important magnetic fields permeating the CSE.

High angular resolution sub-millimetre observations open a new window for studying stellar surface features that may be linked to asymmetry in the mass-loss process. ALMA long baselines (15 km) provide an angular resolution of 25 milli-arcseconds at 230 GHz, making it possible to resolve structures in the stellar disk for some stars. Further out, within a few stellar radii, SiO masers can form and are compact high brightness temperature tracers of the extended atmosphere. Proper motions of the masers provide 3D velocities and indicate outflow and infalling gas in this region, with complex non-radial motion also detected. There is some evidence from different maser species in the circumstellar environment for the presence of dynamically-important magnetic fields, however, at least in the case of SiO masers there are alternative interpretations for the polarization levels detected. ALMA long baseline observations (e.g. Fig.~\ref{fig:humphries}) have also revealed red-shifted molecular absorption close to the star in Mira, providing further evidence of infalling material. ALMA has also been used to study circumstellar dust.

Radio and submm observations support the PEDDRO (Pulsation-Enhanced Dust Driven Outflow, see \citet{Hofner2015}) scenario for evolved stars. Constraints on the propagation of shocks outwards through the (extended) atmosphere are provided by radio photosphere and SiO maser observations. Dust composition and formation radii can be indirectly probed by mapping distributions of dust parent and seed molecules in the gas phase, although care must be taken in interpreting these due to excitation effects. Dust is also studied directly via the sub-millimetre continuum. Wind acceleration can be investigated by studying different water maser lines which straddle the dust formation zone. The role of magnetic fields in AGB and RSG mass loss should be clarified via future multi-wavelength observations.

\begin{figure*}%[!hbtp]
\begin{center}
%\resizebox{\hsize}{!}{
%	\includegraphics[width=0.7\hsize, bb=0 0 504 684, angle=90]{rau.pdf}}
\includegraphics[width=9cm,angle=90]{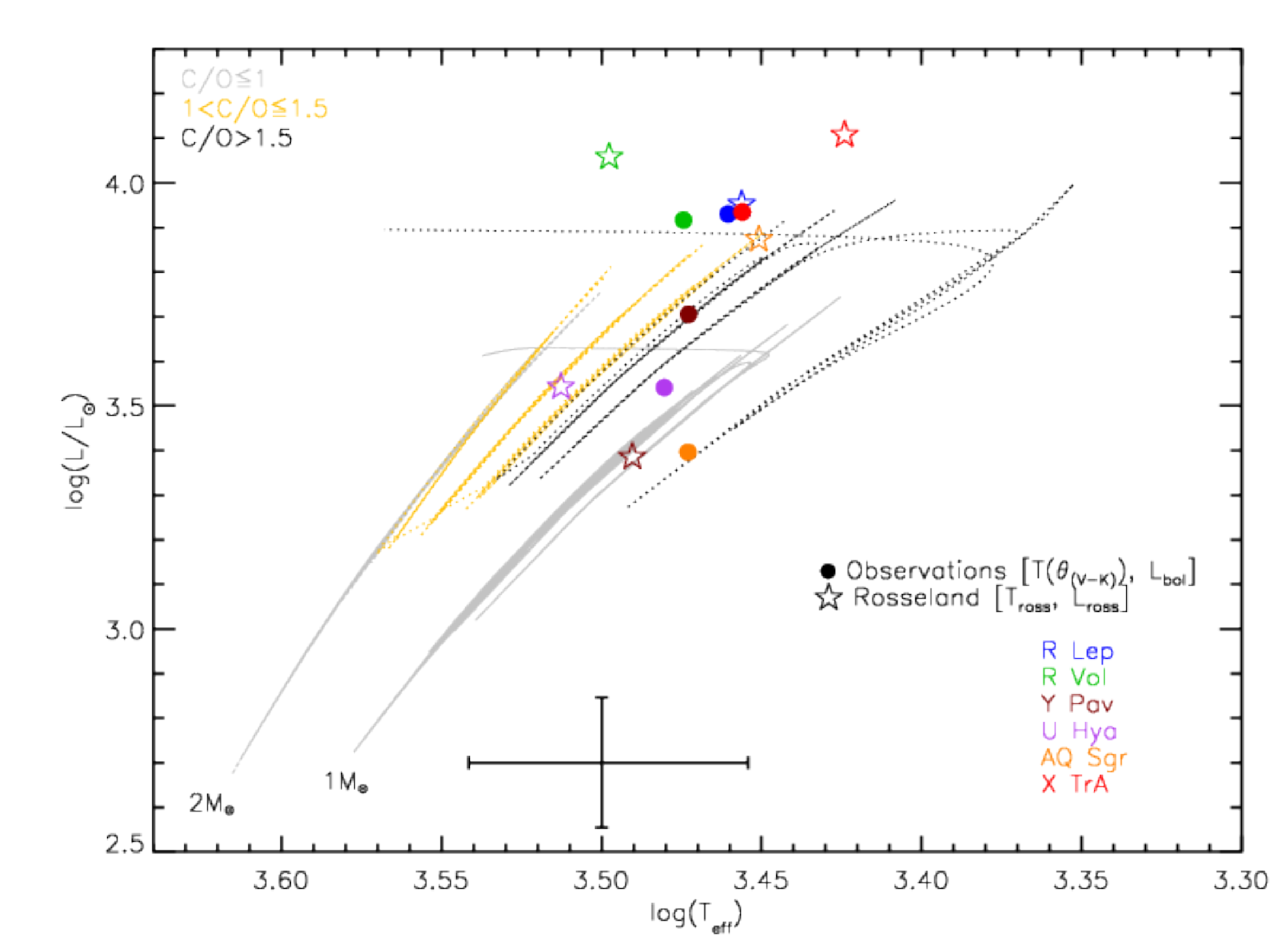}
	\caption{AGB region of the HR diagram (see also Rau et al. (2016 submitted)). The lines show evolutionary tracks with solar metallicity from \cite{marigo13}. The numbers denote the mass values at the beginning of the TP-AGB. A dotted line illustrates the track for $2~$M$_\odot$. Different symbols refer to temperatures and luminosities estimated via observations (solid circles) and via the best fitting models with spectro-photometric-interferometric-observations (open stars). Different colours indicate the various targets. A typical error bar is shown in black in the lower part of the figure.}
	\label{hrdiagram}
\end{center}	
\end{figure*}\
\subsection{Long Baseline Radio Continuum Observations of Betelgeuse:
Understanding Mass Loss in Red Supergiants}
Eamon O'Gorman\footnote{Dublin Institute for Advanced Studies, Republic of Ireland} spoke on behalf of himself,
Pierre Kervella, Leen Decin, Anita Richards, Iain McDonald, Andrea Chiavassa, Xavier Haubois, Guy Perrin, Graham Harper, Miguel Montarg\`es, Keiichi Ohnaka, Jan Martin Winters and Arancha Castro-Carrizo.  He emphasized that spatially resolved cm/mm/sub-mm continuum observations can be a powerful means by which to study the
partially ionized extended atmospheres around red supergiants because they allow us to probe the essential region where mass loss is initiated. To date, only Betelgeuse has been spatially resolved at cm and mm wavelengths
with the data showing that the atmosphere is highly extended and that the mean gas temperature beyond two stellar radii is cool (Tgas < 3000 K). He presented the initial results of their ALMA long baseline and e-MERLIN continuum observations of Betelgeuse's extended atmosphere which probe it at unprecedented angular resolution and sensitivity. The full details will be published elsewhere.

\subsection{Surface magnetic field detection in M type supergiant stars: the non uniqueness of Betelgeuse}
Benjamin Tessore\footnote{LUPM, University of Montpellier, France},  presented work done in collaboration with E.~Josselin, A.~L\`ebre, J.~Morin, A.~L\'opez Ariste, M.~Auri\`ere, Ph.~Mathias and P.~Petit. He suggested that
RSG stars can be considered as the massive counterparts of AGB stars, with which they share properties such as extended atmospheres, prodigious mass loss, and chemically complex circumstellar envelopes.
Magnetic fields have already been measured in their circumstellar envelopes using maser polarization \citep{2014IAUS..302..389V}.

Systematic detections of a surface magnetic field have been reported for a wide sample of giant stars \citep{2013BlgAJ..19...14K,2015A&A...574A..90A}. These measurements have shown that the magnetic giant stars are mostly located in two magnetic strips; one starting at the first dredge-up and one starting at the tip of the RGB phase and along the AGB phase. 
In 2010 \citet{auriere2010} detected a weak magnetic field (at the Gauss level) at the surface of Betelgeuse, prototype of RSG stars. Until recently this was the only red supergiant with a direct detection of a surface field. 

In the spring of 2015 L\`ebre and colleagues started a 2-year campaign with the high resolution (R = 65,000) spectro-polarimeter Narval (TBL Pic du Midi, France) dedicated to a sample of cool evolved stars, including RSG stars such as Betelgeuse.
In polarimetric mode, the Narval instrument simultaneously acquires a classical intensity spectrum as well as a polarized spectrum (either linearly or circularly polarized).

They measured several spectra of three RSG stars, $\alpha$~Her, $\mu$~Cep and CE~Tau, both in circular (Stokes V) and linear polarization (Stokes QU) mode at different epochs.
With the Least-Square Deconvolution (LSD) method \citep{1997MNRAS.291..658D} the spectra of each star are combined to give a mean profile also called an ``LSD profile".
An analysis of these spectra revealed significant hints of a surface field at the Gauss level in these three stars, although in some cases further checks are required to understand fully the nature of the observed polarized signal. They also found  a time variation of these magnetic fields on week/month time scales. This variation is very similar to what is known for Betelgeuse \citep{2013EAS....60..161B, 2013LNP...857..231P} and very consistent with the convective pattern time scales \citep{2002AN....323..213F,2015EAS....71..243M}.

Because of their slow rotation (and thus high Rossby number, up to 100!) convection in these stars is not expected to generate  a global magnetic field efficiently \citep{2004A&A...423.1101D}. Thus these detections may point toward local transitory fields, which could play a role in the mass-loss mechanism.

%\subsubsection{First attempt to map bright spots at the surface of RSG stars}
Recently \citet{Auriere2016} discovered the complex linearly polarized spectrum of Betelgeuse.
This is characterised by structures similar to those found in the classical intensity spectrum. The amplitudes of these structures are well above the typical amplitudes of the Zeeman induced circularly polarized spectrum. Because the amplitude of the Stokes QU LSD profile is between ten and a hundred times greater than the Stokes V profile Auri\`ere {\it et al.} rejected any contribution of the magnetic field to this spectrum. Therefore they have deduced that scattering polarization associated with brightness anisotropies at the surface of Betelgeuse contributes most to this ``second stellar spectrum".

These anisotropies are likely to be caused by the giant convective cells lying at the surface of RSG stars. From an analytical model, and the Stokes QU profiles, they infer the positions of several bright spots at the surface of RSG stars.  Auri\`ere {\it et al.} have thus mapped two bright spots at the surface of Betelgeuse. Remarkably the latter results are in good agreement with interferometric models of \citet{2015EAS....71..243M}.

%\subsubsection{Future Prospects}

Although magnetic fields are a key ingredient of many evolution codes they are still poorly understood. Moreover, as with the Sun, the linearly polarized spectrum, which seems to be a common features of RSG stars, has opened a new window for diagnostics of their surface dynamics. Tessore and colleagues have therefore  undertaken a long-term monitoring campaign of RSG stars. They also plan to extend the interpretation of the linear polarization to other cool evolved stars, such as Miras and RV~Tauri stars, where again rather strong linearly polarized spectra are detected.

\begin{figure*}
\centering
\includegraphics[width=12cm]{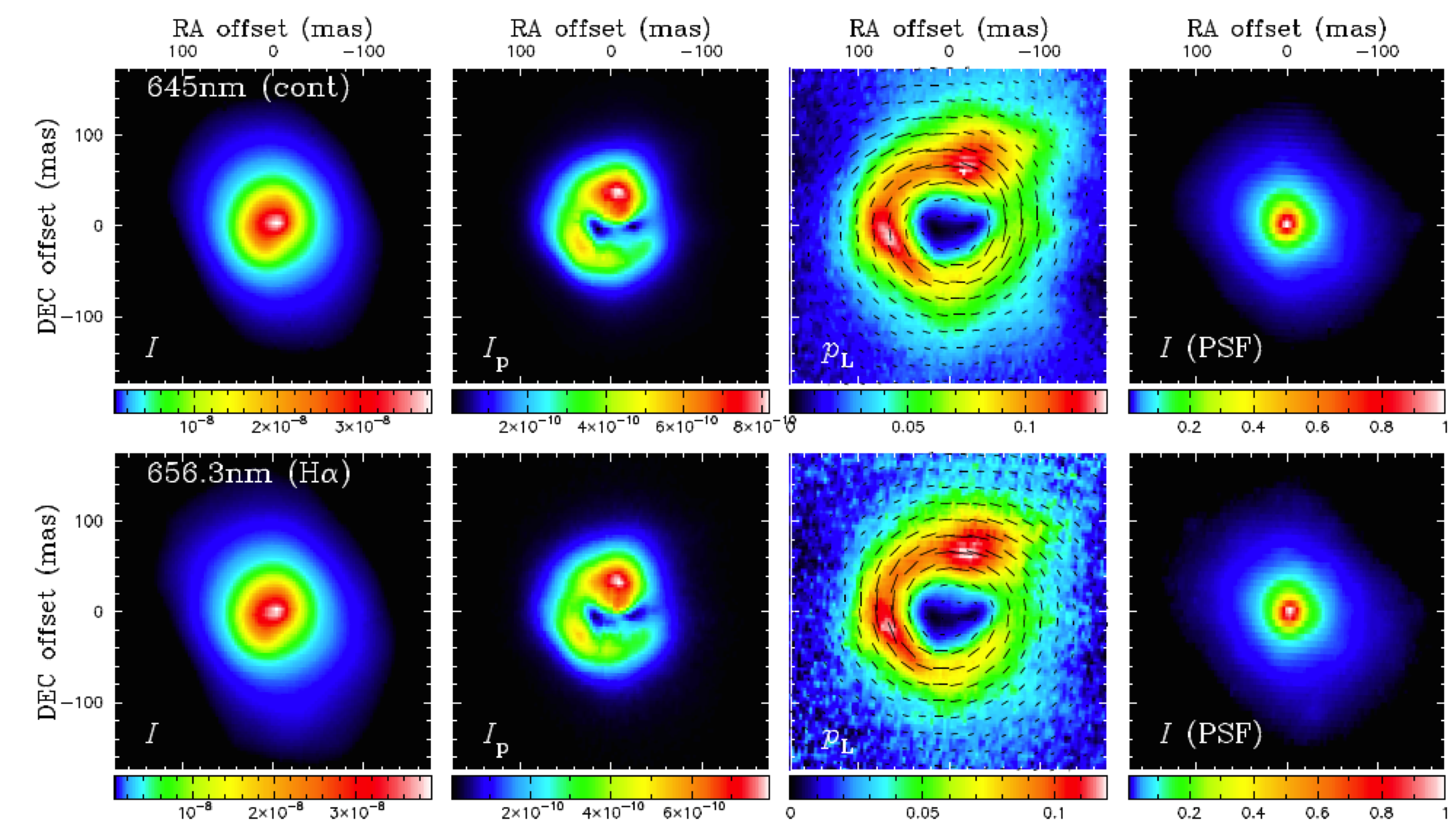}
   \caption{SPHERE/VLT observations of W Hya (Ohnaka et al. 2016). }
    \label{fig:eric1}
\end{figure*}

\subsection{Modelling a set of carbon-rich AGB stars at high-angular resolution}
\label{hron}
Joseph Hron\footnote{University of Vienna, Austria} gave this presentation which had been prepared by his student Gioia Rau in collaboration with himself, C. Paladini, B. Aringer, K. Eriksson, P. Marigo, W. Nowotny and R. Grellmann.

They compared spectrophotometric and VLTI/MIDI interferometric observations of six carbon-rich AGB stars, R~Lep, R~Vol, Y~Pav, AQ~Sgr, U~Hya and X~TrA, with self-consistent dynamic models atmospheres from the Uppsala group \citep{Eriksson2014,Mattsson2010}. The results were somewhat similar to those in  \cite{rau15}.   The models reproduce the SED data well at wavelengths longward of $1~\mu$m, and the interferometric observations between $8~\mu$m and $10~\mu$m. Shortwards of $1~\mu$m in the SED, and longwards of $10~\mu$m in the visibilities, they found discrepancies which may be due to a combination of data- and model-related effects.

The models which fit the Miras best are considerably extended, with a significant shell-like structure. In contrast, the models which fit the non-Miras best are more compact and show lower average mass-loss rates. They derived the following stellar parameters from the fits: effective temperature ($T_{\rm eff}$), Luminosity ($L_{\rm models}$), Mass ($M$), $C/O$ and mass-loss rate ($\dot{M}$). The results are, within the uncertainties, in good agreement with values from the literature. $T_{\rm eff}$ agrees with the temperature derived from the angular diameter $\theta_{(V-K)}$ and the bolometric luminosity from the SED fitting, $L_{bol}$, except for AQ~Sgr. 

Finally, the Rosseland diameter $\theta_{Ross}$ and $\theta_{(V-K)}$ \citep{vanbelle2013} agree with each other better for the Mira targets than for the non-Miras, possibly because of episodic mass loss of the latter models.

%The estimated $T_\text{eff}$, $L_\text{eff}$ are used to place the stars on the HR diagram. 
The stars can be compared to evolutionary tracks in an HR diagram (see Fig.~\ref{hrdiagram}). The main  properties (L, T$_{\rm eff}$, C/O ratios and stellar masses) derived from the model fitting result in good agreement with thermal pulsing (TP)-AGB evolutionary calculations for carbon stars \citep{marigo13}. %Those were carried out with the COLIBRI code.

The detailed findings will be presented by Rau et al. (2016 submitted to A\&A).

\section{Day 2: The circumstellar environment of cool giants}
Evolved stars eject much of their mass into space through extreme winds. The ejecta form an
expanding shell around the star, extending from the dust formation radius at two stellar radii to the
asteropause at 1-4 parsec where the merger with the interstellar medium takes place. Although the
winds are thought to be largely spherically symmetric, recent observations show growing evidence
for strong shaping. The ESO VLT and ALMA have found evidence for spirals, circumstellar disks,
bipolar flows and jets, and even wind-blown tails. The dominant shaping mechanisms is believed
to be angular momentum, which for most of stellar evolution is in cold storage in stellar, sub-stellar
and planetary companions, but is transferred to the shells through dynamical interactions. Related
processes act in supergiants and their descendant supernovae, while common envelope systems and stellar mergers,
cover most of the HR diagram. We brought together observers and theoreticians to discuss dust production and 
shaping mechanisms over a wide range of stellar properties. 
%Topics include the new observations at extreme angular resolution (e.g. Sphere, ALMA), 3D hydrodynamical models, planetary systems around evolved stars, and common envelope evolution.

\begin{figure}
\centering
\includegraphics[width=8cm]{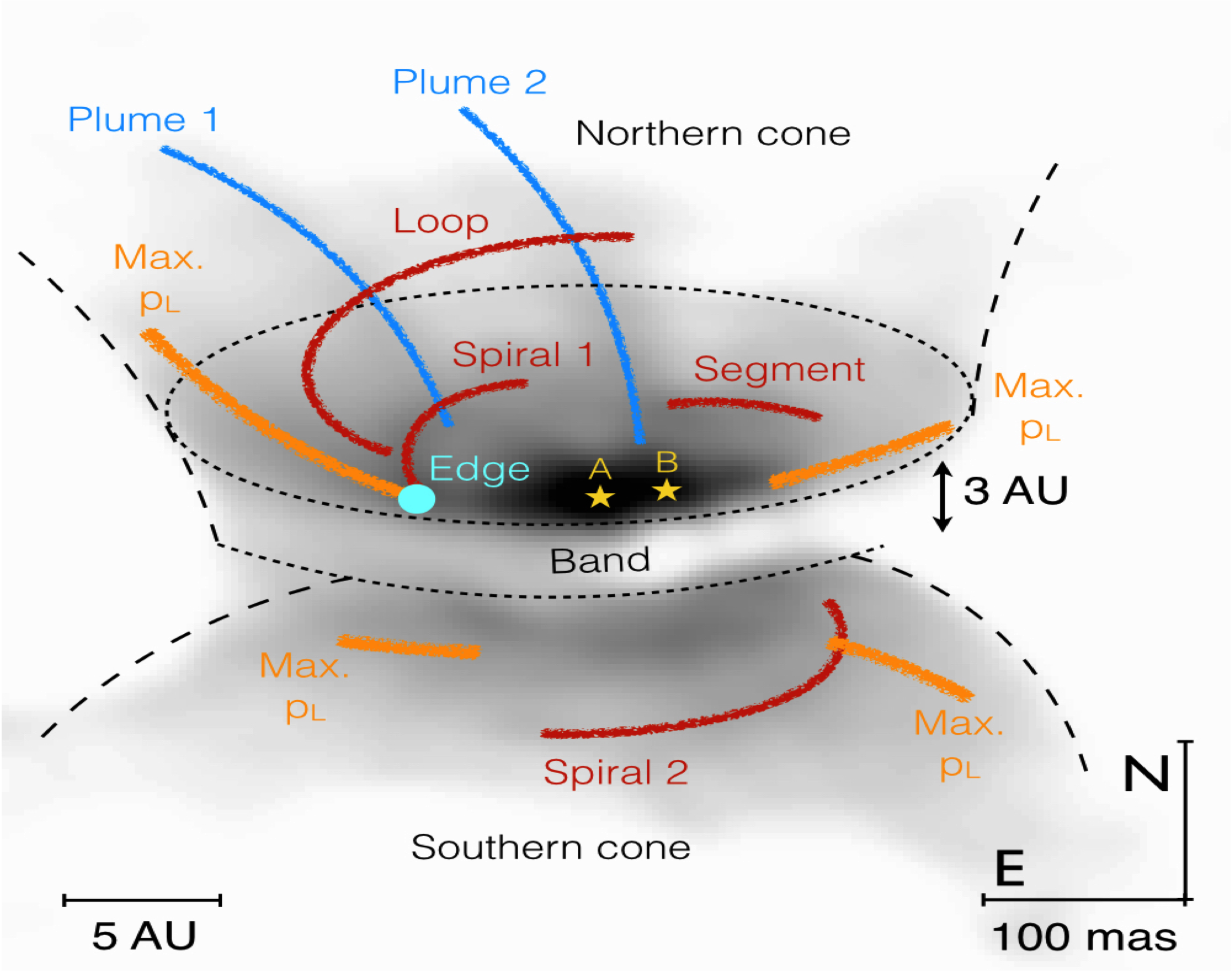}
   \caption{SPHERE/VLT observations of L2 Pup showing the disc with ejected plumes \citep{Kervella2015}. }
    \label{fig:eric2}
\end{figure}

\subsection{The circumstellar environment of AGB and RSG stars}
Eric Lagadec\footnote{Observatoire de la Cote d'Azur, France} started his invited review by highlighting the fact that new high-angular-resolution instruments such as ALMA, SPHERE/VLT and VLTI/PIONIER have started a small revolution in the study of the circumstellar environments around evolved stars. We no longer talk about 
sub-arcsecond observations, but about milli-arcsec observations. This is revolutionising our view of the close environment of these stars, the way they interact with binary companions and our understanding of the mass-loss process.

Mass loss from evolved stars is a key process for the final stages of evolved stars, as it determines how these stars will enrich the interstellar medium in newly synthesised elements and  thus the contribution of stars to the chemical enrichment of galaxies. For mass loss to occur there is a need to first levitate the atmosphere and then accelerate it so that it can escape the gravitational attraction of the star. This is described by Susanne H\"ofner as a two-stage rocket mechanism.
For decades, it was thought that the mass-loss process from AGB  stars was fully understood. Convection and pulsation trigger shocks, leading to the formation of dust behind these shocks, where the gas is cool and dense. Radiation pressure on dust grains then initiates the mass loss and the gas is carried along via collisions. A study by \citet{Woitke2006} revealed that this could not work for oxygen-rich AGB stars, as the silicate grains forming around these stars cannot provide the opacity needed for radiation pressure to work.
 %(Woitke P., 2006, A\&A, 460,9).
 \citet{Hofner2008} thus suggested that the mass loss from these stars was due to scattering rather than absorption.
For scattering to be efficient large grains ($\sim 1 \mu m$) are needed.
RSG stars also lose mass and their winds certainly share many similarities with AGB stars. RSGs have larger masses than AGBs, but lower pulsation amplitudes and smaller more numerous convective cells than their low mass analogues. Models predict that the atmospheres of AGB stars are more extended than those of RSGs.

Knowing these theoretical aspects of AGB and RSG  stars, it is  interesting to see what high angular resolution observations can tell us about them. Optical/infrared interferometry can teach us about the extension of the atmospheres, the gas dynamics, convection and dust formation. Extreme adaptive optics systems, such as SPHERE/VLT, when combined with polarimetric measurements can tell us about the dust distribution, grain size and convection. Performing time series  observations around the pulsation/convection cycle allows us to link convection and pulsation with shocks and dust formation. %He briefly reviewed these techniques and highlighted a few key results.

Long-baseline infrared interferometry enables us to reach resolutions down to one mas with instruments such as AMBER/VLTI, PIONIER/VLTI, and thus even to map the surface of a few nearby stars. Combining light from more than two telescopes brings imaging capabilities and can provide resolved images of convective cells for nearby stars like Betelgeuse \citep{Haubois2009}. Combined with high spectral resolution, interferometry can help mapping the MOLSPHERE \citep [see e.g.][]{Ohnaka2016} and study the gas dynamics by mapping the gas distribution at different wavelengths. Such observations reveal that for AGB and RSG stars, the atmosphere is more extended in the CO and water bands than in the continuum and that the extension of the atmosphere is similar in AGB and RSG stars \citep{Wittkowski2016, Arroyo2015}. This is not reproduced by RSG models suggesting that an extra mechanism is necessary to explain the mass-loss mechanism of RSGs. This could be radiation pressure on lines \citep{Josselin2007}. High spectral resolution observations by \citet{Ohnaka2014} reveal that the atmosphere of the RSG Antares appears different across the CO line profile and that the velocity field is not homogeneous. Changes can be seen within a year. All this indicates that we are observing the evolution of convective cells at the surface of Antares.

\begin{figure*}
\centering
\includegraphics[width=16cm]{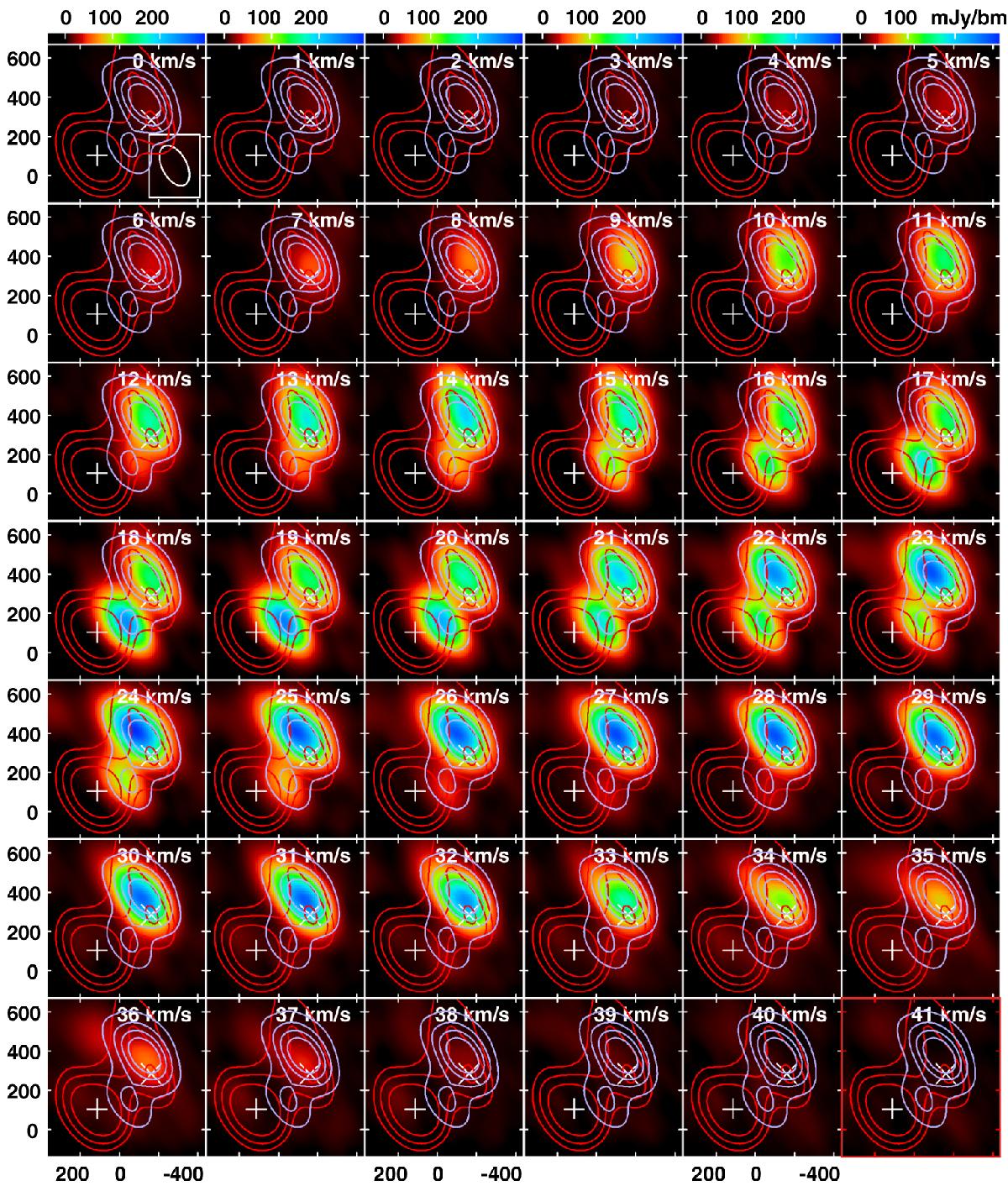}
  \caption{ALMA channel maps (colour scale) of the NaCl v=0 J=24-23 emission at 312\,GHz at a 1\,km s$^{-1}$ velocity resolution for the central $0".8\times0".8$ field around the stellar position of the red supergiant VY~CMa. Grey contours are the integrated NaCl line strength at $(1, 2, 3, 4)\times 1$\,Jy/beam km\,s$^{-1}$, red contours are the dust continuum measured with ALMA at 321\,GHz at $(1, 2, 3, 4) \times 18$\,mJy/beam \citep{Richards2014, Gorman2015}. The stellar position `VY' is indicated with a white cross, and the peak of the continuum emission `C' is indicated with a white plus-sign. The abscissa and ordinate show the offset of the right ascension and declination, respectively, in units of milli-arcseconds \citep{decin2016}.}
    \label{fig:decin}
\end{figure*}

As mentioned above, extreme adaptive optics combined with polarimetry allows us to study the dust distribution and grain size.
\citet{Ohnaka2016} obtained spectacular results from observations of the AGB star W~Hya with SPHERE/VLT (Fig.~\ref{fig:eric1}). They conclude that the polarimetric signal from dust in the envelope of the AGB star W Hya requires the presence of large grains (0.4-0.5 $\mu$m). This is consistent with the predictions of models proposing that mass loss is driven by scattering. A few large dust clouds are also observed, as expected when large-scale convective flows help levitate the atmosphere. Similar results were obtained with SPHERE/VLT for the AGB star  R~Dor, and the  surface of the star was even resolved \citep{Khouri2016}, showing variation within less than two months. SPHERE/VLT observations of the RSG Betelgeuse also enable the surface to be mapped \citep{Kervella2016}, showing the presence of asymmetries likely due to convection and the presence of dust close to the star. All these results indicate that the mass loss from both O-rich AGB and RSG stars is very likely linked to convection and scattering by large dust grains close to the surface of the stars.

Another striking aspect of the new high-angular-resolution observations of AGBs and RSGs is that large asymmetries are observed, e.g. in VY~CMa \citep{Gorman2015} and the Fried Egg Nebula \citep{Wallstrom2015}. Some extra momentum (the third stage of the rocket) is needed to explain these structures and could come from a binary companion and/or a magnetic field.
Such morphologies and momentum excesses have been observed in many planetary nebulae. In a key paper \citet{Bujarrabal2001} showed that a large majority of proto-planetary nebulae have outflow momenta in excess of that supplied by radiation pressure. A long debate took place in the planetary nebulae community and two key papers helped settle the matter. \citet{Soker2004} showed that a single star cannot supply enough energy and angular momentum to shape those nebulae. In the meantime, \citet{Nordhaus2006} demonstrated that magnetic fields can play an important role, although isolated stars cannot sustain a magnetic field for long enough. Looking for direct and indirect evidence for binaries became a goal of the field. Soon, observations by Olivier Chesneau and his team revealed that discs, formed via the  interaction between the evolved star and a companion, were common \citep{Chesneau2006, Chesneau2007}.  An important joint effort to look for binaries (using radial velocities or infrared excess techniques) led to the conclusion that binaries were common and shaping the nebulae \citep[see e.g.][]{Boffin2012}.
It is clear that if binaries play a key role in the shaping of planetary nebulae they must  play a role in the previous phase of evolution, the AGB. Unfortunately, AGB stars pulsate and are bright infrared emitters, so the techniques used to find binary companions around Planetary Nebulae cannot be used.

High angular resolution mapping of AGB stars has, however, been shown to be an excellent tool to find binaries (in a direct or indirect way). One of the key recent results for the studies of these binary companions was the   combination of a theoretical work led by Shazrene Mohamed and ALMA observations by Maercker and collaborators. \citet{Mohamed2007} showed that the interaction of an AGB star with a  companion could lead to Wind Roche-lobe overflow and the formation of 3D spiral structures. Such a spiral was found with ALMA observation of R~Scl \citep{Maercker2012}.
Looking for spirals thus became a technique to find the binary companions of AGB and RSG stars. A spiral was found around Mira \citep{Ramstedt2014}, IRC+10216 \citep{Decin2015} and more are being discovered. Binaries thus seem to play a key role in the shaping of evolved stellar envelopes.

Finally, SPHERE/VLT's first observations of an AGB star offered us a very spectacular view of  L2~Pup (Kervella et al., 2015, Fig.~\ref{fig:eric2}). At 64\,kpc this is one of the closest AGB stars, enabling us to obtain sharp views of its structure. The SPHERE/VLT images directly revealed the presence of an edge-on disc, with plumes of material being ejected perpendicular to it, as predicted by theoretical models for the formation of bipolar nebulae via the interaction with binary companions. These observations revealed a secondary source at 2\,AU. Have we mapped for the first time an AGB star with a binary companion and  a dusty disc formed via the interaction of the binary system? (a future bipolar planetary nebula?) This will have to be confirmed via radial velocity and ALMA's ongoing monitoring of the system.

In conclusion: the new generation of high angular resolution  instruments is revolutionising our view of evolved stars and making this an exciting time to study such objects.  There are strong indications that convection and scattering via dust play a key role in the mass loss from RSG and O-rich AGB stars. For the first time, we have mapped directly the surfaces of stars other than the Sun and seen them evolving over a few months. It also turns out that binaries are rather common among evolved stars and influence the shape of the circumstellar environment. 

To improve our understanding of mass loss from these stars, we need time-series observations to study quantitatively  the physical processes such as shocks and dust formation. To  understand quantitatively the impact of AGB and RSG stars on the chemical enrichment of galaxies via the mass-loss process we need to measure  the physical properties (mass-loss rates, binary influence etc.) for a statistically significant sample. We are currently focusing on a few spectacular targets, but we need to understand the whole stellar population. This should help reach one of our goals: being able to determine the dust and gas mass-loss rates from evolved stars knowing only their fundamental parameters (mass, age, binarity, etc.).

\begin{figure}
        \centering
        \includegraphics[angle=180,width=0.48\textwidth]{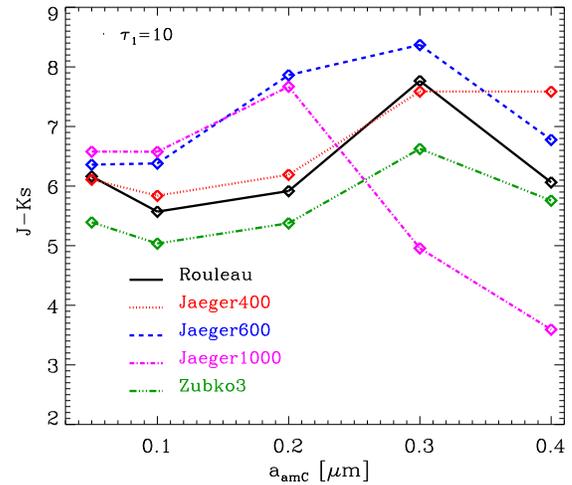}
        \caption{$J-K_S$ colour as a function of the grain size for selected data sets of amorphous carbon dust. The models are computed for $\tau_1=10$.}
        \label{nanni1}
        \end{figure}
        
        \begin{figure}
        \centering
\includegraphics[angle=180, width=0.48\textwidth]{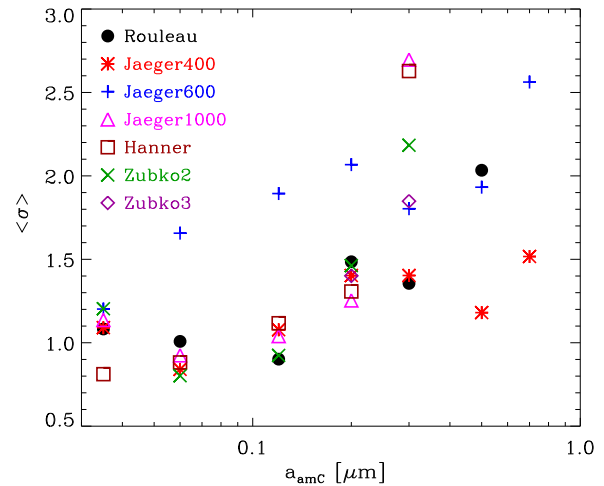}
        \caption{Deviations between models and observations as a function of the grain size, for different combinations optical data sets available in the literature.}
        \label{nanni2}
        \end{figure}

 \begin{figure*}
 \centering
  \includegraphics[angle=180, width=0.48\textwidth]{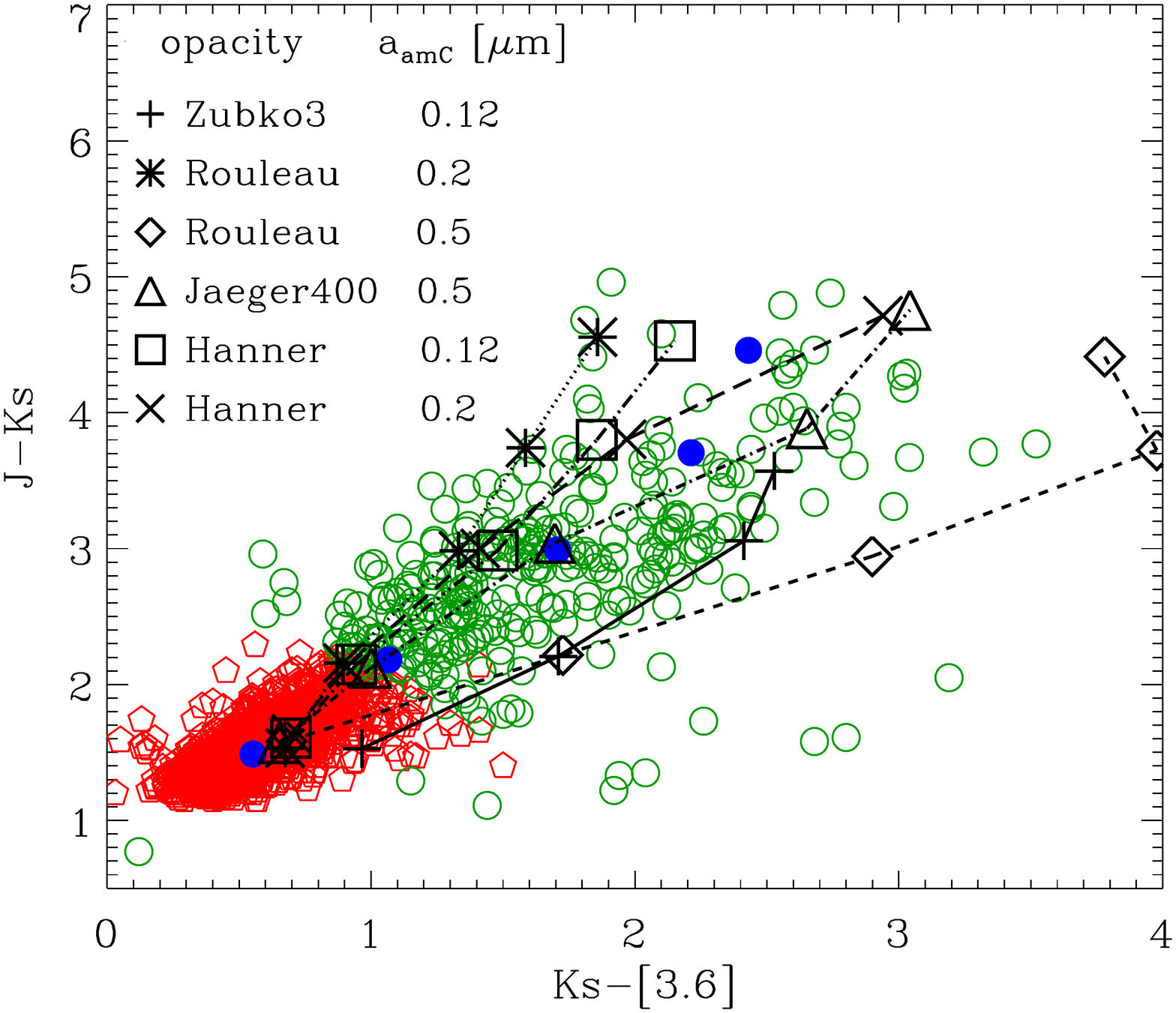}
   \includegraphics[angle=180, width=0.48\textwidth]{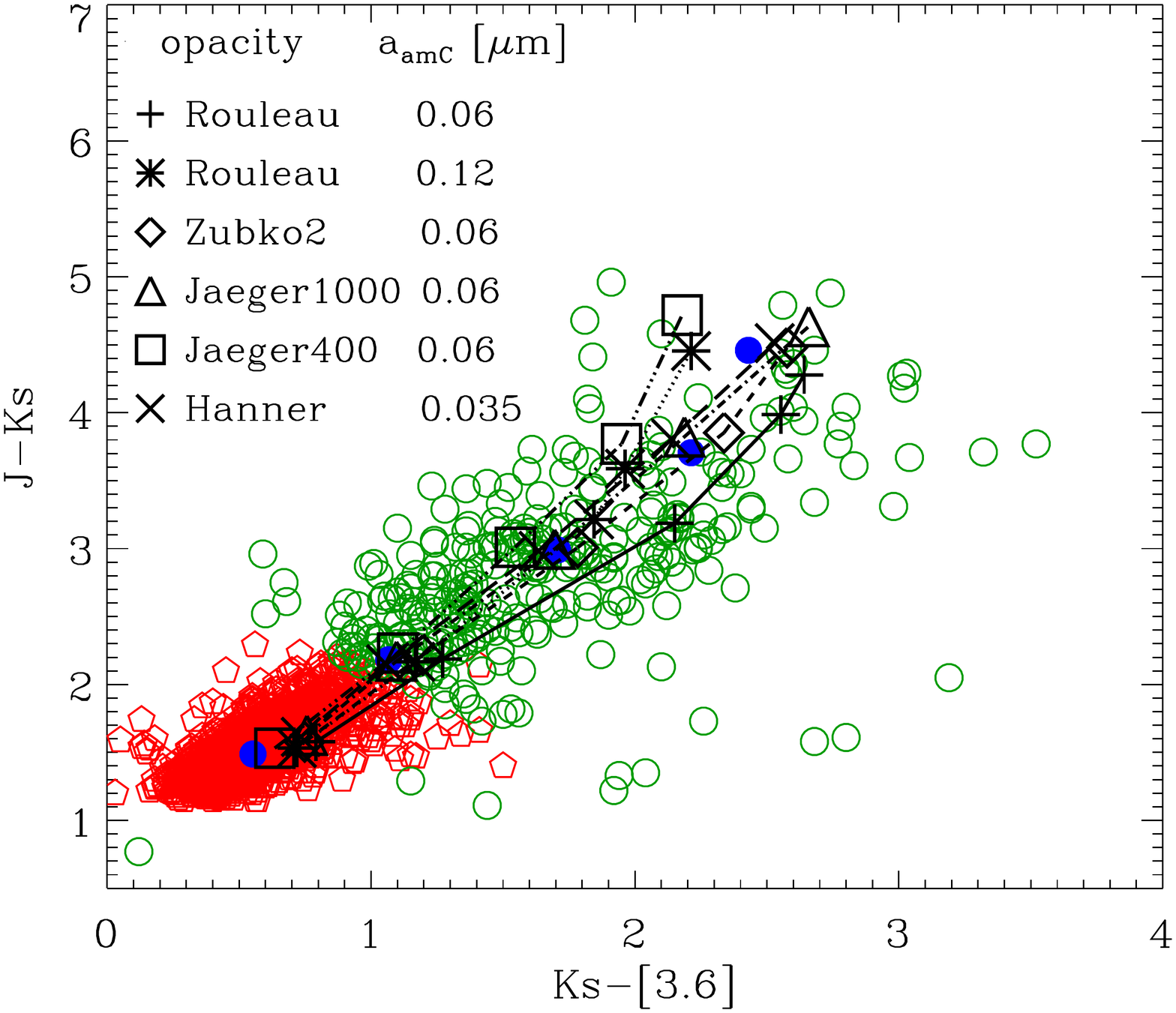}
        \caption{$J-K_S$ vs $K_S-[3.6]$ colour-colour diagrams for the observed sample of C- (red pentagons) and extreme-stars (open green circles) superimposed with the average value for models obtained with different optical data sets and grain size, listed in the top left. Full blue circles represent the average values of the observed stars in different $J-K_S$ intervals. In the left panel some combinations of optical data  and grain sizes which poorly reproduce the observations are shown, while in the right panel some of the well performing combinations are plotted.}
        \label{nanni3}
        \end{figure*}

\subsection{High spatial resolution studies of the winds of evolved stars} 
Leen Decin\footnote{Institute of Astronomy, Leuven University, Belgium} started by highlighting the amazing view we now have of the winds of evolved cool
giants and supergiants provided by high spatial resolution observations. We can  see that the long-standing assumption of smooth spherically symmetric winds is very often completely invalid. 
Large scale structures in the form of spirals, circumstellar disks, bipolar outflows, bowshocks etc.\ are detected and smaller scale clumpiness seems omnipresent (Fig.~\ref{fig:decin} illustrates NaCl observations of VY~CMa). These novel data challenge our understanding of the wind launching process. The observations serve as critical benchmarks for 3D hydrodynamical models.
Even more, these data push the theoretical models to include a higher form of complexity, in particular to incorporate a (more) consistent approach to chemistry, dynamics, and radiative transfer. In her talk she reviewed the recent observational results obtained with ALMA, SPHERE and Herschel. She went on to discuss
how these data yield detailed information on the wind structure of evolved stars and elucidate which chemical
and physical phenomena should be captured in theoretical wind models. 

She also summarized ongoing efforts to improve these theoretical models, both in terms of numerical modeling and based on novel laboratory
experiments. In turn these theoretical models can serve as a guideline to further improve the observing strategies.

\subsection{Constraining dust properties in Circumstellar Envelopes of C-stars in the Small Magellanic Cloud: optical constants and grain size of carbon dust} 
Ambra Nanni\footnote{Dipartimento di Fisica e Astronomia ``G. Galilei'', Padova, Italy} presented on behalf of herself and Paola Marigo, Martin Groenewegen, Berhard Aringer, L\'eo Girardi, Giada Pastorelli, Alessandro Bressan and Sara Bladh. She talked about recent investigation aimed at constraining the typical size and optical properties of carbon dust grains in circumstellar envelopes (CSEs) of C-stars in the SMC.

They applied their recent dust growth model, coupled with a radiative transfer code, to the CSEs of C-stars evolving along the TP-AGB, for which they compute spectra and colours. Then they compare the modelled colours with those observed in the near-infrared (NIR) and mid-infrared (MIR) bands, testing different assumptions about the dust and employing optical constants from various sources for carbon dust.  Different assumptions within the dust model change the typical size of the carbon grains produced.
They finally constrain carbon dust properties by selecting the combination of grain sizes and optical constants which best reproduces several colours in the NIR and MIR at the same time.
The different choices of optical properties and grain size can lead to differences
in the NIR and MIR colours greater than two magnitudes. As an example of this, Fig.~\ref{nanni1} illustrates the $J-K_S$ colour computed for different optical data sets as a function on the grain size for a dust-enshrouded CSE with $\tau_1=10$.

It seems that the complete set of selected NIR and MIR colours are best reproduced by small grains, with sizes between 0.06 and 0.1 $\mu$m, rather than by large grains of 0.2-0.4 $\mu$m. Remarkably, the inability of large grains to reproduce NIR and MIR colours seems independent of the adopted optical data set and the deviations between models and observations tend to increase for increasing grain sizes. The trend recovered is shown in Fig.~\ref{nanni2}, where the deviations between models and observations, $<\sigma>$, are plotted as a function of the grain size and for different optical data sets.
An example of  a colour-colour diagram is shown in Fig.~\ref{nanni3}, which illustrates the colours of the  stars observed superimposed on models for poorly performing combinations of optical constants and grain sizes (left panel) and for well performing ones (right panel).
There is also a possible trend of the typical grain size with the dust reddening in the CSEs of these stars.

She emphasized that this work is preparatory to follow-up studies aimed at calibrating the
TP-AGB phase through resolved stellar populations in star clusters and galaxies which include dusty, mass-losing evolved stars.

\subsection{The Dustiest AGB Stars and RSG in the Magellanic Clouds}

Olivia Jones\footnote{STScI, Baltimore, USA}  described using observations from the {Herschel Inventory of The Agents of Galaxy Evolution (HERITAGE) survey of the Magellanic Clouds \citep{Meixner2013, Seale2014}, enabling them to identify 32 evolved stars that are bright in the far-infrared. These extremely red late-type stars span a wide range of initial masses, luminosities, dust temperatures and chemical type.

They found 13 low- to intermediate-mass evolved stars, AGB stars, post-AGB stars, planetary nebulae and a symbiotic star. They also identify 10 high mass stars, including three extreme red supergiants that are highly enshrouded by dust, and detect nine probable evolved objects which had not previously been described in the literature. These sources are among the dustiest evolved objects in the Magellanic Clouds. The Herschel emission may either be due to dust produced by the evolved star or it may arise from material swept-up from the interstellar medium (ISM) \citep{Jones2015}.

The detection of these evolved stars in the far-infrared may have far reaching consequences for global dust-budget estimates as we do not currently account for all their large dust reservoirs. Thus the global dust return to the ISM may have been significantly underestimated.

%Several of these stars are not currently accounted for in dust budget estimates for the Large Magellanic Cloud. If this dust is produced by the evolved star during the current epoch of mass-loss, then the total dust return from AGB stars and RSGs in the LMC could increase by $\sim$30--50\%.

Figure~\ref{fig:LMCSED_eg} presents the spectral energy distributions for a selection of AGB stars and RSGs with a far-infrared emission in the LMC; Fig.~\ref{fig:hessCMDirac} shows their location in CMD space.

\begin{figure*} 
\centering
\includegraphics[width=15cm]{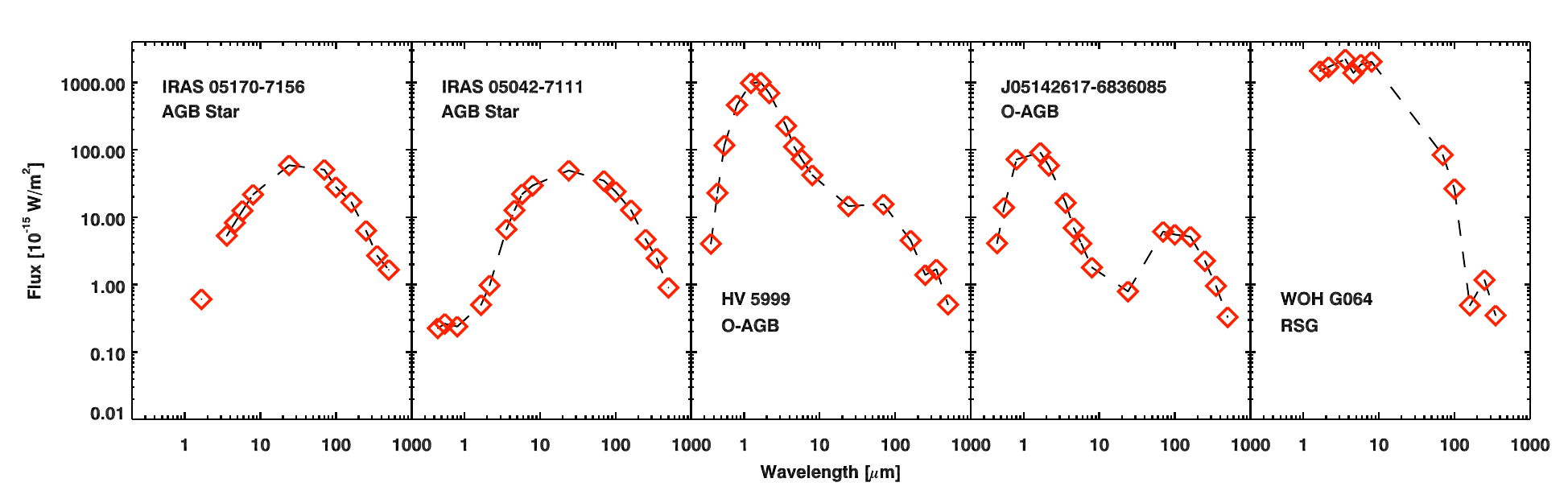}
%\vspace{0.3cm}
 \caption[LMC SEDs]{The spectral energy distributions for a selection of AGB stars and RSGs with a far-infrared emission in the LMC. The red diamonds are the near-to far-infrared $UBVI$, $JHK$, Spitzer and Herschel photometry.} 
  \label{fig:LMCSED_eg}
\end{figure*}

\subsection{Modelling the interactions of cool giants with (sub)stellar companions: 
Implications for mass-transfer rates and outflow geometries}

Shazrene Mohamed\footnote{South African Astronomical Observatory, Cape Town, South Africa} presented an invited paper on behalf of herself, 
Ph.\ Podsiadlowski, R.\ Booth, E.\ Aydi, M.\ Maercker, S.\ Ramstedt,  and W.\ Vlemmings.
Cool evolved stars, e.g., AGB and RSG stars, lose copious amounts of mass and momentum through powerful dense stellar winds. The interaction of these outflows with their surroundings results in highly structured and complex circumstellar environments, often featuring knots, arcs, shells and spirals. Recent improvements in computational power and techniques have led to the development of detailed multi-dimensional simulations that have given new insight into the origin of these structures, and better understanding of the physical mechanisms driving their formation.                        	

One of the main mechanisms for shaping the outflows of evolved stars is interaction with a companion. The primary structure that forms is a spiral outflow due to the orbital motion of the giant and the orbit and gravitational focusing of the companion. The inter-arm spacing between the windings depends primarily on the orbital period and wind velocity. The contrast between the emission from the arm and inter-arm regions also depends on the mass of the companion. Thus, by comparing models and observations of these structures we can constrain, not only the orbital characteristics, but also the evolution of the stellar wind properties, e.g., mass-loss rate and velocity (as was done for R~Scl, Maercker et al. 2012). More recently, they have received new ALMA data for the known binary system, Mira~AB. Many of the circumstellar structures fit well with expectations from the Wind Roche-Lobe Overflow (WRLOF) models of \citet{Mohamed2012} (see Fig.~\ref{mohamed}). There are, however, structures, e.g., the heart-shaped bubble/`racoon' which are likely the result of the interaction between Mira's wind and an outflow from the companion \citep{Ramstedt2014}.

\begin{figure} 
\centering
\includegraphics[width=3.5in]{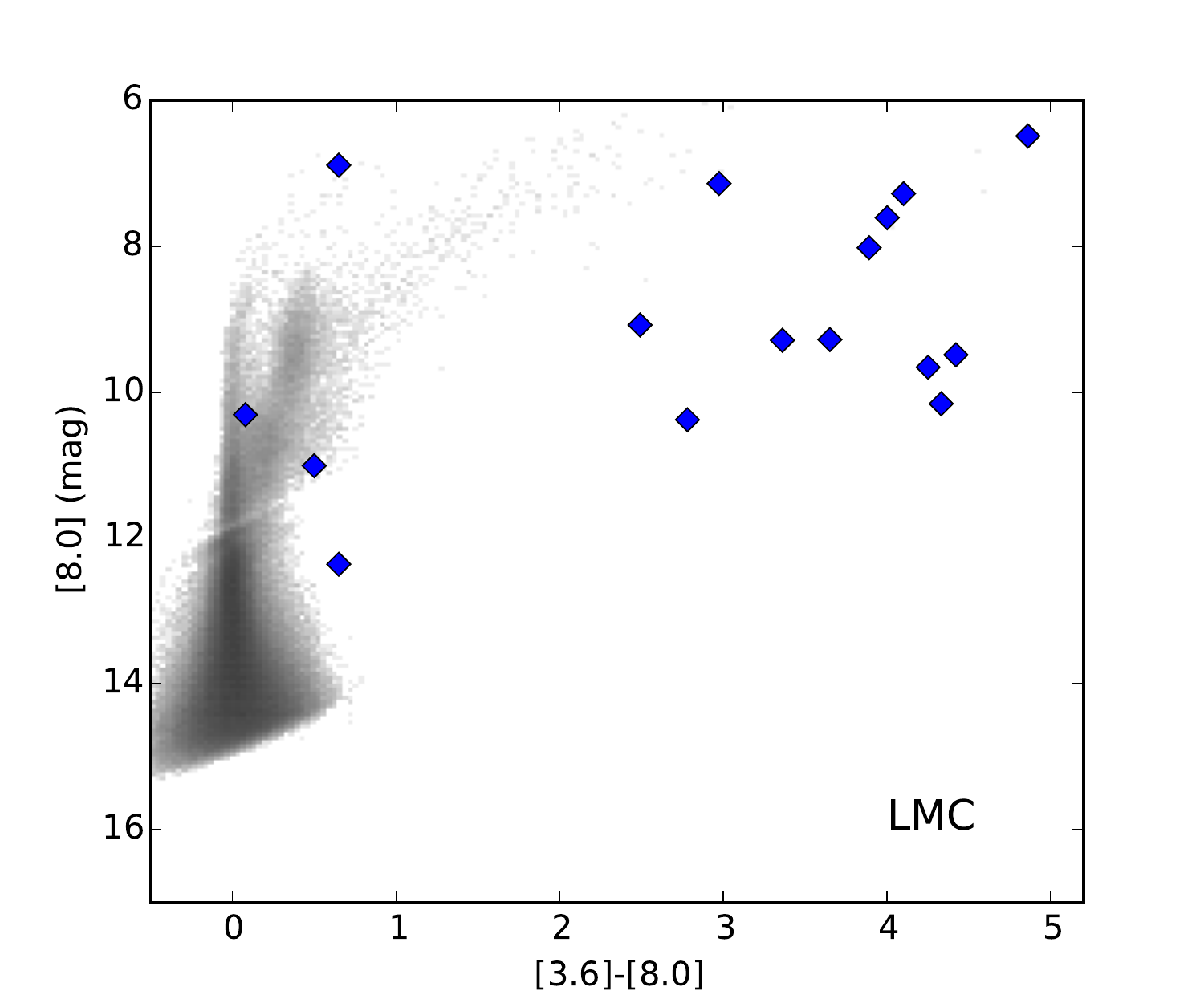}
 \caption{The [8.0] vs. [3.6]--[8.0] CMD for the LMC showing the evolved stars with far-infrared emission (blue diamonds) overlaid on a Hess diagram of the evolved stars identified by \cite{Boyer2011} from the SAGE-LMC survey \citep{Meixner2006}.} 
  \label{fig:hessCMDirac}
\end{figure}

Indeed, in many cool giant binary systems, not only does the companion affect the outflow, but the outflow can have important implications for the companion. In systems where material is accreted the companion may increase in mass, and produce high energy radiation and outflows of its own (both via steady winds and explosive outbursts). These binary interactions thus have broad implications for a wide range of phenomena, e.g., the formation of planetary nebulae, symbiotic and X-ray binaries, novae and supernovae. RS~Oph, a recurrent nova system, is a particularly interesting example as the companion is a massive white dwarf, making the system a potential Type Ia supernova progenitor. The system also shows Na\,{\sc i}\,D line variations which result from the interaction of nova ejecta and radiation with the dense circumstellar outflow from the giant; similar variations are seen in about 20\% of Type Ia supernovae \citep{Patat2007,Patat2011,Simon2009,Sternberg2011}. Mohamed's models of the binary interaction and recurrent novae explosions fit the overall observed morphology \citep{OBrien2006} and demonstrate the dependence of the Na\,{\sc i}\,D line variations on the viewing angle \citep{Booth2016}. 

\begin{figure*} 
\centering
\includegraphics[width=12cm]{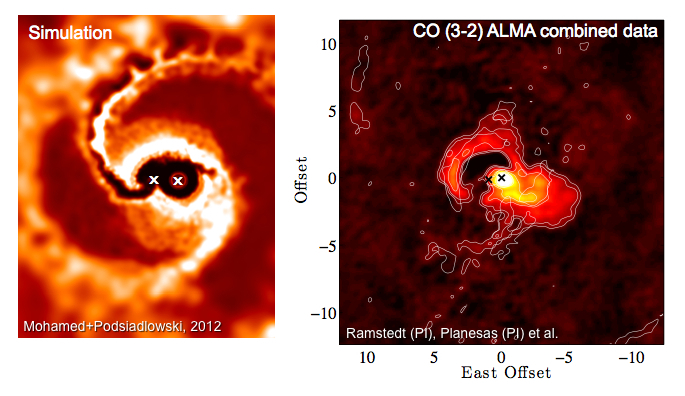}
 \caption{Left: WRLOF simulation - Slice in the binary orbital plane showing the dust opacity that results from the interaction of a giant star with a Mira-type outflow (including pulsations and radiation pressure on dust) and a $0.6\, M_{\odot}$ companion. 
Right: ALMA CO (3-2) observation of the binary star system, Mira AB.} 
  \label{mohamed}
\end{figure*}
\subsection{Dust Production Rates of AGB Stars in the Solar Neighbourhood}
Alfonso Trejo-Cruz\footnote{Institute of Astronomy \& Astrophysics, Academia Sinica, Taiwan} reminded us that AGB stars are a very important contributor to the total dust mass injected into the ISM in galaxies. Good estimations of the dust mass injection by these stars in the Magellanic Clouds have been achieved using Spitzer data \citep{Riebel2012, Srinivasan2016}. However, the last estimate of the dust injection rate in the Milky Way was made in the late '80s for a limited sample \citep{Jura1989}. So Trejo-Cruz and colleagues decided to revisit the total dust mass-loss rate from AGB stars in the Solar neighborhood. 

It is especially hard to evaluate distances to dusty AGB stars in the Milky Way, as the highest mass-loss rate objects are not in the Hipparcos catalogue, due to circumstellar extinction. Using present-day all-sky infrared surveys (WISE, 2MASS, and others) they constructed spectral energy distributions for all AGB stars within one kiloparsec of the Sun. They use the GRAMS model grid \citep{Sargent2011, Srinivasan2011} to estimate the dust production for this sample of AGB stars. Preliminary results show an increase in the number of known dusty objects within one kpc. An integrated dust production rate of 
$\sim$ 10$^{-5}$ M$_{\odot}$ yr$^{-1}$ or an average of $\sim$ 10$^{-7}$ M$_{\odot}$ yr$^{-1}$ per object is obtained. These results for the Solar Neighborhood will be extrapolated for the entire Milky Way, using a suitable stellar distribution function. They compared these  results to those of the Magellanic Clouds and other Local Group galaxies, for which the distance determination problems do not exist. This work is a step towards more reliable determinations of the mass loss rates of AGB stars, and it aims to provide new insight on the discrepancy between the dust mass produced by AGB stars and that estimated to be present in the ISM.

\subsection{A First Look at the Circumstellar Dust Composition of M-type Mira Variables observed with Spitzer}
Tina G\"uth\footnote{New Mexico Institute of Mining and Technology, New Mexico Tech., USA} spoke, on behalf of herself and Michelle  Creech-Eakman, about the details of the composition of the dust surrounding Mira variables. These regular pulsators are easily identified in the optical and infrared due to their changes in brightness. Data on 25 Galactic Miras were obtained with the Spitzer Infrared Spectrograph (IRS) instrument in 2008-09 under a GO program \citep{Creech-Eakman2008}. IRS has four different modules to allow for detailed observations: SL, LL, SH, and LH. The stars were observed once per month to track changes in their brightness and spectral features. This data set is unique both for the number of observations of each star and for the high SNR made possible because of the stars' intrinsic brightness. Short exposure times were used to avoid saturation of the detectors. 

The current focus of their research is the short, high (SH) and long, high (LH) resolution wavelength range ($\sim9.7$ $ - 40 \mu$m) because it was least likely to saturate the detectors. The reduction process for the high resolution data was roughly based on recipes from the Spitzer Data Analysis  Cookbook. It should be noted that each observation had a dedicated sky background taken, as recommended by the Spitzer Space Telescope team due to low cryogen levels and potential problems at the end of the cryogenic mission. In the data reduction process, the first step is the subtraction of the dedicated sky background from the data instead of the more commonly used nod-nod subtraction. The next steps involved bad pixel removal (IRSCLEAN), spectrum extraction (SPICE tool), defringing (IRSFRINGE), order overlap removal (Spitzer Data Analysis Cookbook), and nod combination (IRAF and own IDL codes). Fig.~\ref{guth} is an example of final spectra obtained.

\begin{figure*} 
\centering
\includegraphics[width=14cm]{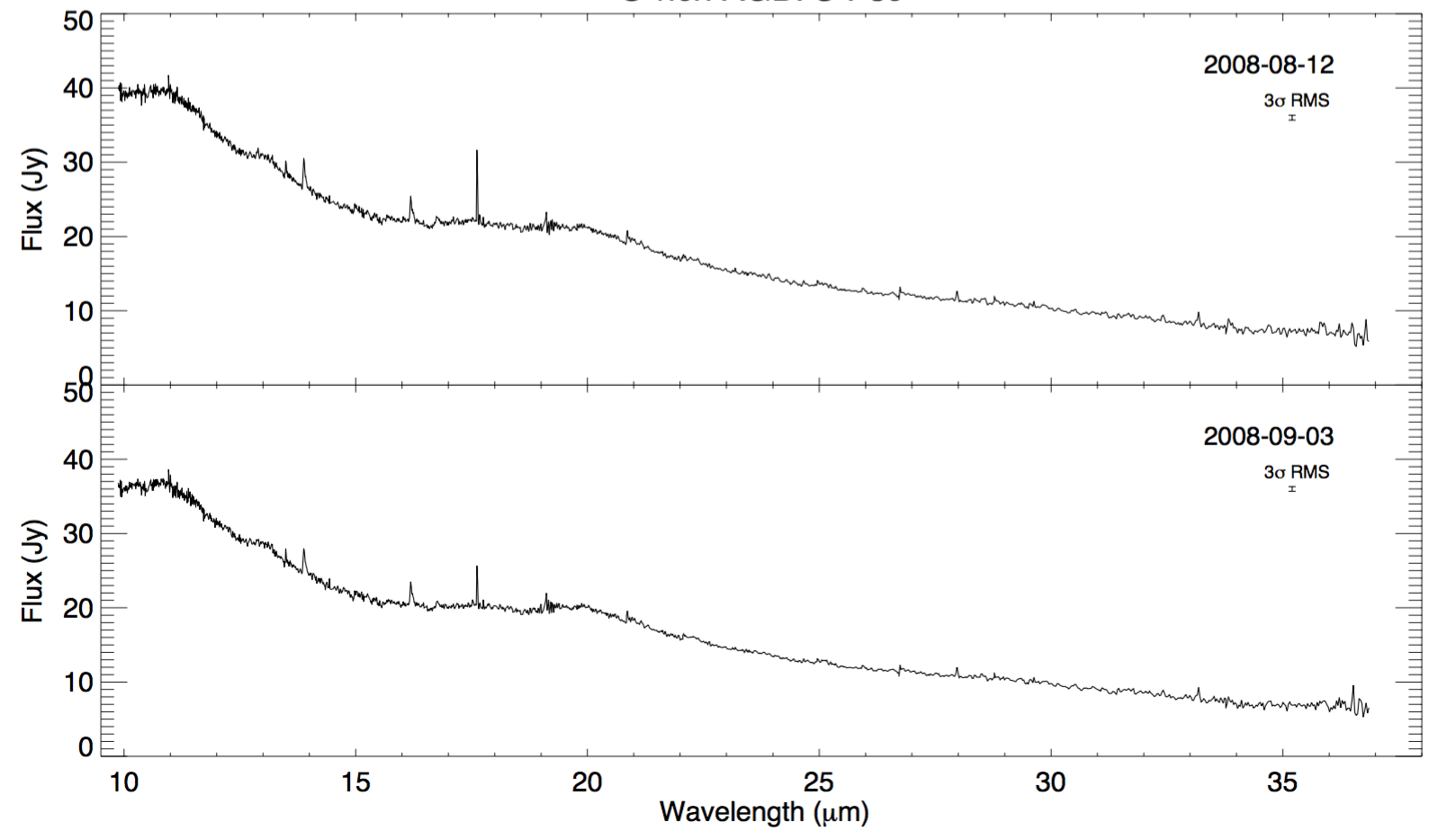}
%\vspace{0.3cm}
 \caption{Spitzer spectra of the O-rich AGB star S Per.} 
  \label{guth}
\end{figure*}

\begin{figure*} 
\centering
\includegraphics[width=14cm]{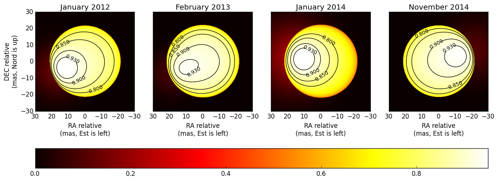}
%\vspace{0.3cm}
 \caption{Intensity maps of Betelgeuse derived from the best fitted LDD and Gaussian hotspot model on the VLTI/PIONIER data. North is up and East to the left.} 
  \label{mont1}
\end{figure*}

The high resolution, reduced spectra reveal a wonderful ``forest'' of features that provide insight into the stellar atmospheres and circumstellar dust composition. The stars in this study span the range of oxygen- to carbon-rich, with each type exhibiting certain known solid state components (i.e.\ dust) as previously identified using IRAS LRS and ISO LWS \citep[e.g.,][]{Little-Marenin1990, Tsuji1997}. Preliminary examination of the oxygen-rich Miras shows many of the primary known features, such as broad silicate emission and aluminum oxide ($\rm Al_2O_3$). The radiative transfer modeling code, DUSTY, is being used to identify several broad, and some sharp, dust features, after implementing recently derived laboratory spectral indices for dust opacities. Obtaining dust opacity data in the mid-infrared range for elements and compounds other than those provided with DUSTY has been difficult due to the limited publication of the  index information necessary for use with DUSTY. Some opacity tables have been provided by Dr. Angela Speck (University of Missouri). However, more data are needed to obtain the detailed circumstellar dust composition of these Miras.

Prominent features seen in oxygen-rich Mira variables include initial identifications of water ice emission, as well as amorphous and crystalline silicates, and potentially corundum. However, the spectrum from DUSTY does not align well with the Spitzer spectrum across the entire flux range, which makes some feature identification difficult. Dr. Bernhard Aringer (University of Padova/STARKEY) recommended the use of a greybody continuum rather than a blackbody continuum to overcome this problem. In particular, Dr. Aringer and Dr. Claudia Paladini (Universit\'e Libre de Bruxelles) both suggest the MARCS code. They are currently in the process of implementing these recommendations to obtain better fits and improve identification of the dust and other features in the spectra.
\begin{figure*} 
\centering
\includegraphics[width=14cm]{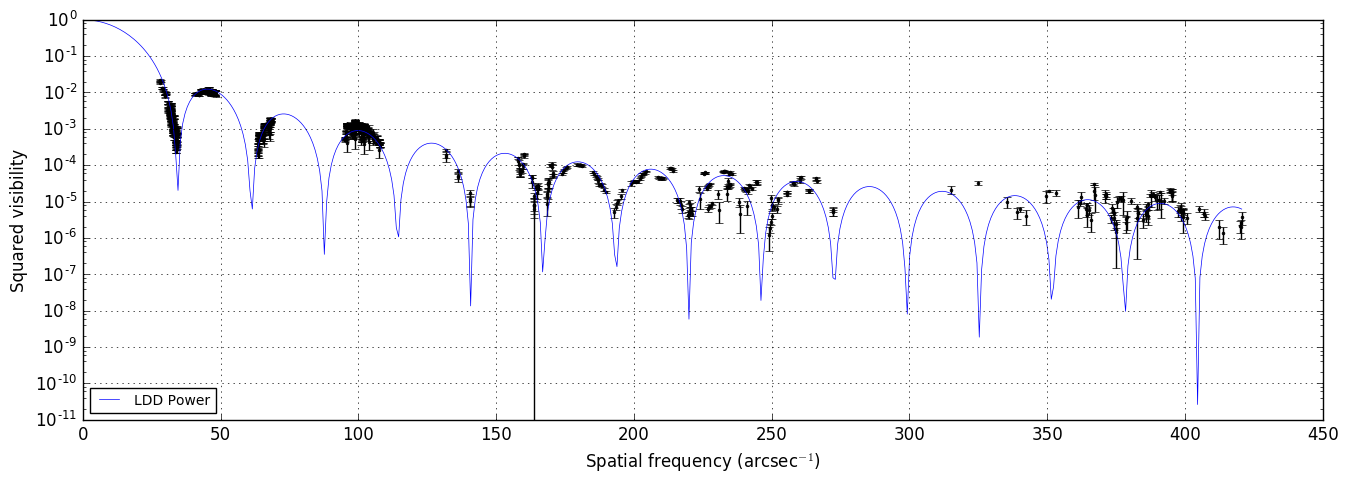}
%\vspace{0.3cm}
 \caption{Squared visibilities measured by PIONIER on Antares (black). The blue continuous line represents the best fit limb-darkened disk (power-law from \citet{Hestroffer1997}).} 
  \label{mont2}
\end{figure*}

\subsection{The optical interferometry view of convection in nearby red supergiants}
Miguel Montarg\`es\footnote{IRAM, Grenoble, France} spoke on behalf of himself, P.\ Kervella, G.\ Perrin, A.\ Chiavassa, M.\ Auri\`ere, A.\ L\'opez-Ariste and P.\ Mathias, on the subject of RSG stars. The huge increase in size of RSG stars when they leave the main sequence at a relatively constant mass causes their surface gravity to be quite weak and may be linked to their important mass loss. However, the physical processes which effectively trigger the outflow are not fully understood.  RSGs are among the stars with the largest apparent diameter in our sky. Therefore long baseline optical interferometry makes it possible to resolve the photosphere from where the mass loss originates. 

 Montarg\`es and collaborators obtained VLTI/PIONIER observations of the two nearby RSG stars: Betelgeuse ($\alpha$~Ori) and Antares ($\alpha$~Sco). PIONIER is a near-infrared instrument that recombines the light of four telescopes of the Paranal Observatory operated by ESO. It operates in the $H$-band and they used only the high spectral resolution mode (R~40).

	The observations of Betelgeuse sampled four epochs between January 2012 and November 2014, using only the compact configuration of the 1.8m Auxiliary Telescopes (AT).  These observations show that the first zero of the visibility function is located at a spatial frequency that depends on the direction sampled on the (u, v) plane. 
Together with the strong deviations of the closure phase signal from 0$^{\rm o}$ and 180$^{\rm o}$ this indicates the presence of strong asymmetries on the stellar surface. The four epochs of observation are well fitted by a power-law limb-darkened disk (LDD) plus a Gaussian hot spot. The intensity images resulting from fitting this model are presented in Fig.~\ref{mont1} \citep{Montarges2016}.  They interpreted this hot spot as the bright top of a large convective cell. They compared their interferometric measurements with 3D radiative hydrodynamics simulations \citep{Chiavassa2011}, but the convective patterns do not seem to match the data, even if they did so in the past. The currently available simulations seem  unsuited to reproduce the present convective regime of Betelgeuse induced by the presence of this large bright convective cell. The existence of such features is confirmed by the TBL/Narval spectro-polarimetric observations of \citet{Auriere2016}, (see also Tessore's talk in this splinter session). VLT/SPHERE observations indicate the presence of dust at three stellar radii of Betelgeuse \citep[][and Lagadec's talk]{Kervella2016} in the same direction (North-East) as the bright spot  Montarg\`es observed. Additionally, ALMA observations show an asymmetry also in the same direction (O'Gorman's talk). The connection between the optical and millimetre interferometric observations and the SPHERE data may be directional mass loss, of convective or of some other nature.

	In May 2014 they observed Antares using the same instrument with the same setup. This time they obtained data with the three available AT configurations, with baselines ranging from 11 to 156m. Thus they were able to probe the visibility function up to the 16th lobe (an angular resolution of ~3 mas = 1/15th of the stellar diameter). These unprecedented data will put new constraints on the numerical simulations by resolving the mean size of convective features. In particular, they detected excesses with respect to the limb-darkened disk (Fig.~\ref{mont2}) at 100 arcsec$^{-1}$ (12 mas),  200-270 arcsec$^{-1}$ (4.5-5.6 mas) and 350-400 arcsec$^{-1}$ (3-3.5 mas).

	These high angular resolution observations of two bright nearby RSG stars in the near-infrared with long baseline interferometry allowed us to better characterize the convective pattern of these stars. This is essential to uncover the link between convection and  mass loss in RSG.

\section*{Acknowledgments}
{We are grateful to all of the speakers at the splinter session and to the following who provided written material and/or figures for this summary paper: Claudia Paladini, Sofie Liljegren, Sara Bladh, Liz Humphreys, Benjamin Tessore, Gioia Rau, Eric Lagadec, Ambra Nanni, Olivia Jones, Shazrene Mohamed, Alfonso Trejo-Cruz, Tina G\"uth and Miguel Montarg\`es.
Patricia Whitelock thanks the National Research Foundation of South Africa for a research grant. }

\bibliographystyle{cs19proc}
\bibliography{merged.bib}

\begin{thebibliography}{101}
\providecommand{\natexlab}[1]{#1}

\bibitem[\protect\astroncite{{Arroyo-Torres}
  \emph{et~al.}}{2014}]{arroyo-torres2014}
{Arroyo-Torres}, B., {Mart{\'{\i}}-Vidal}, I., {Marcaide}, J.~M., {Wittkowski},
  M., {Guirado}, J.~C., \emph{et~al.} 2014, \aap, 566, A88.

\bibitem[\protect\astroncite{{Arroyo-Torres} \emph{et~al.}}{2015}]{Arroyo2015}
{Arroyo-Torres}, B., {Wittkowski}, M., {Chiavassa}, A., {Scholz}, M.,
  {Freytag}, B., \emph{et~al.} 2015, \aap, 575, A50.

\bibitem[\protect\astroncite{{Arroyo-Torres}
  \emph{et~al.}}{2013}]{arroyo-torres2013}
{Arroyo-Torres}, B., {Wittkowski}, M., {Marcaide}, J.~M., \& {Hauschildt},
  P.~H. 2013, \aap, 554, A76.

\bibitem[\protect\astroncite{{Auri{\`e}re} \emph{et~al.}}{2010}]{auriere2010}
{Auri{\`e}re}, M., {Donati}, J.-F., {Konstantinova-Antova}, R., {Perrin}, G.,
  {Petit}, P., \emph{et~al.} 2010, \aap, 516, L2.

\bibitem[\protect\astroncite{{Auri{\`e}re}
  \emph{et~al.}}{2015}]{2015A&A...574A..90A}
{Auri{\`e}re}, M., {Konstantinova-Antova}, R., {Charbonnel}, C., {Wade}, G.~A.,
  {Tsvetkova}, S., \emph{et~al.} 2015, \aap, 574, A90.

\bibitem[\protect\astroncite{{Auri{\`e}re} \emph{et~al.}}{2016}]{Auriere2016}
{Auri{\`e}re}, M., {L{\'o}pez Ariste}, A., {Mathias}, P., {L{\`e}bre}, A.,
  {Josselin}, E., \emph{et~al.} 2016, \aap, 591, A119.

\bibitem[\protect\astroncite{{Barnbaum} \& {Hinkle}}{1995}]{barnbaum1995}
{Barnbaum}, C. \& {Hinkle}, K.~H. 1995, \aj, 110, 805.

\bibitem[\protect\astroncite{{Baron} \emph{et~al.}}{2014}]{baron2014}
{Baron}, F., {Monnier}, J.~D., {Kiss}, L.~L., {Neilson}, H.~R., {Zhao}, M.,
  \emph{et~al.} 2014, \apj, 785, 46.

\bibitem[\protect\astroncite{{Bedecarrax}
  \emph{et~al.}}{2013}]{2013EAS....60..161B}
{Bedecarrax}, I., {Petit}, P., {Auri{\`e}re}, M., {Grunhut}, J., {Wade}, G.,
  \emph{et~al.} 2013, In \emph{EAS Publications Series}, edited by
  P.~{Kervella}, T.~{Le Bertre}, \& G.~{Perrin}, \emph{EAS Publications
  Series}, vol.~60, pp. 161--165.

\bibitem[\protect\astroncite{{Bladh} \emph{et~al.}}{2015}]{Bladh2015}
{Bladh}, S., {H{\"o}fner}, S., {Aringer}, B., \& {Eriksson}, K. 2015, \aap,
  575, A105.

\bibitem[\protect\astroncite{{Boffin} \emph{et~al.}}{2012}]{Boffin2012}
{Boffin}, H.~M.~J., {Miszalski}, B., {Rauch}, T., {Jones}, D., {Corradi},
  R.~L.~M., \emph{et~al.} 2012, Science, 338, 773.

\bibitem[\protect\astroncite{{Booth} \emph{et~al.}}{2016}]{Booth2016}
{Booth}, R.~A., {Mohamed}, S., \& {Podsiadlowski}, P. 2016, \mnras, 457, 822.

\bibitem[\protect\astroncite{{Boyer} \emph{et~al.}}{2011}]{Boyer2011}
{Boyer}, M.~L., {Srinivasan}, S., {van Loon}, J.~T., {McDonald}, I., {Meixner},
  M., \emph{et~al.} 2011, \aj, 142, 103.

\bibitem[\protect\astroncite{{Bujarrabal} \emph{et~al.}}{2001}]{Bujarrabal2001}
{Bujarrabal}, V., {Castro-Carrizo}, A., {Alcolea}, J., \& {S{\'a}nchez
  Contreras}, C. 2001, \aap, 377, 868.

\bibitem[\protect\astroncite{{Chandler} \emph{et~al.}}{2007}]{chandler2007}
{Chandler}, A.~A., {Tatebe}, K., {Hale}, D.~D.~S., \& {Townes}, C.~H. 2007,
  \apj, 657, 1042.

\bibitem[\protect\astroncite{{Chesneau} \emph{et~al.}}{2006}]{Chesneau2006}
{Chesneau}, O., {Collioud}, A., {De Marco}, O., {Wolf}, S., {Lagadec}, E.,
  \emph{et~al.} 2006, \aap, 455, 1009.

\bibitem[\protect\astroncite{{Chesneau} \emph{et~al.}}{2007}]{Chesneau2007}
{Chesneau}, O., {Lykou}, F., {Balick}, B., {Lagadec}, E., {Matsuura}, M.,
  \emph{et~al.} 2007, \aap, 473, L29.

\bibitem[\protect\astroncite{{Chiavassa} \& {Freytag}}{2015}]{chiavassa2015}
{Chiavassa}, A. \& {Freytag}, B. 2015, In \emph{Why Galaxies Care about AGB
  Stars III: A Closer Look in Space and Time}, edited by F.~{Kerschbaum}, R.~F.
  {Wing}, \& J.~{Hron}, \emph{Astronomical Society of the Pacific Conference
  Series}, vol. 497, p.~11.

\bibitem[\protect\astroncite{{Chiavassa} \emph{et~al.}}{2011}]{Chiavassa2011}
{Chiavassa}, A., {Freytag}, B., {Masseron}, T., \& {Plez}, B. 2011, \aap, 535,
  A22.

\bibitem[\protect\astroncite{{Chiavassa}
  \emph{et~al.}}{2010{\natexlab{a}}}]{chiavassa2010b}
{Chiavassa}, A., {Haubois}, X., {Young}, J.~S., {Plez}, B., {Josselin}, E.,
  \emph{et~al.} 2010{\natexlab{a}}, \aap, 515, A12.

\bibitem[\protect\astroncite{{Chiavassa}
  \emph{et~al.}}{2010{\natexlab{b}}}]{chiavassa2010a}
{Chiavassa}, A., {Lacour}, S., {Millour}, F., {Driebe}, T., {Wittkowski}, M.,
  \emph{et~al.} 2010{\natexlab{b}}, \aap, 511, A51.

\bibitem[\protect\astroncite{{Creech-Eakman}
  \emph{et~al.}}{2008}]{Creech-Eakman2008}
{Creech-Eakman}, M., {Hora}, J., {Ivezic}, Z., {Jurgenson}, C., {Luttermoser},
  D., \emph{et~al.} 2008, \emph{{An Interferometric Snapshot Survey to
  Constrain Mass-Loss Dynamics and Physics in AGB Stars}}.
\newblock Spitzer Proposal.

\bibitem[\protect\astroncite{{Cruzal{\`e}bes}
  \emph{et~al.}}{2015}]{cruzalebes2015}
{Cruzal{\`e}bes}, P., {Jorissen}, A., {Chiavassa}, A., {Paladini}, C.,
  {Rabbia}, Y., \emph{et~al.} 2015, \mnras, 446, 3277.

\bibitem[\protect\astroncite{{Decin} \emph{et~al.}}{2016}]{decin2016}
{Decin}, L., {Richards}, A.~M.~S., {Millar}, T.~J., {Baudry}, A., {De Beck},
  E., \emph{et~al.} 2016, \aap, 592, A76.

\bibitem[\protect\astroncite{{Decin} \emph{et~al.}}{2015}]{Decin2015}
{Decin}, L., {Richards}, A.~M.~S., {Neufeld}, D., {Steffen}, W., {Melnick}, G.,
  \emph{et~al.} 2015, \aap, 574, A5.

\bibitem[\protect\astroncite{{Deroo} \emph{et~al.}}{2007}]{deroo2007}
{Deroo}, P., {van Winckel}, H., {Verhoelst}, T., {Min}, M., {Reyniers}, M.,
  \emph{et~al.} 2007, \aap, 467, 1093.

\bibitem[\protect\astroncite{{Donati}
  \emph{et~al.}}{1997}]{1997MNRAS.291..658D}
{Donati}, J.-F., {Semel}, M., {Carter}, B.~D., {Rees}, D.~E., \& {Collier
  Cameron}, A. 1997, \mnras, 291, 658.

\bibitem[\protect\astroncite{{Dorch}}{2004}]{2004A&A...423.1101D}
{Dorch}, S.~B.~F. 2004, \aap, 423, 1101.

\bibitem[\protect\astroncite{{Eriksson} \emph{et~al.}}{2014}]{Eriksson2014}
{Eriksson}, K., {Nowotny}, W., {H{\"o}fner}, S., {Aringer}, B., \& {Wachter},
  A. 2014, \aap, 566, A95.

\bibitem[\protect\astroncite{{Feast} \emph{et~al.}}{1989}]{Feast1989}
{Feast}, M.~W., {Glass}, I.~S., {Whitelock}, P.~A., \& {Catchpole}, R.~M. 1989,
  \mnras, 241, 375.

\bibitem[\protect\astroncite{{Fox} \& {Wood}}{1982}]{FoxWood1982}
{Fox}, M.~W. \& {Wood}, P.~R. 1982, \apj, 259, 198.

\bibitem[\protect\astroncite{{Freytag} \& {H{\"o}fner}}{2008}]{freytag2008}
{Freytag}, B. \& {H{\"o}fner}, S. 2008, \aap, 483, 571.

\bibitem[\protect\astroncite{{Freytag}
  \emph{et~al.}}{2002}]{2002AN....323..213F}
{Freytag}, B., {Steffen}, M., \& {Dorch}, B. 2002, Astronomische Nachrichten,
  323, 213.

\bibitem[\protect\astroncite{{Haubois} \emph{et~al.}}{2009}]{Haubois2009}
{Haubois}, X., {Perrin}, G., {Lacour}, S., {Verhoelst}, T., {Meimon}, S.,
  \emph{et~al.} 2009, \aap, 508, 923.

\bibitem[\protect\astroncite{{Haubois} \emph{et~al.}}{2015}]{haubois2015}
{Haubois}, X., {Wittkowski}, M., {Perrin}, G., {Kervella}, P., {M{\'e}rand},
  A., \emph{et~al.} 2015, \aap, 582, A71.

\bibitem[\protect\astroncite{{Hestroffer}}{1997}]{Hestroffer1997}
{Hestroffer}, D. 1997, \aap, 327, 199.

\bibitem[\protect\astroncite{{Hofmann} \emph{et~al.}}{2014}]{Hofmann2014}
{Hofmann}, K.-H., {Weigelt}, G., \& {Schertl}, D. 2014, \aap, 565, A48.

\bibitem[\protect\astroncite{{H{\"o}fner}}{2008}]{Hofner2008}
{H{\"o}fner}, S. 2008, \aap, 491, L1.

\bibitem[\protect\astroncite{{H{\"o}fner}}{2015}]{Hofner2015}
{H{\"o}fner}, S. 2015, In \emph{Why Galaxies Care about AGB Stars III: A Closer
  Look in Space and Time}, edited by F.~{Kerschbaum}, R.~F. {Wing}, \&
  J.~{Hron}, \emph{Astronomical Society of the Pacific Conference Series}, vol.
  497, p. 333.

\bibitem[\protect\astroncite{{H{\"o}fner} \emph{et~al.}}{2016}]{Hofner2016}
{H{\"o}fner}, S., {Bladh}, S., {Aringer}, B., \& {Ahuja}, R. 2016, ArXiv
  e-prints.

\bibitem[\protect\astroncite{{Jones} \emph{et~al.}}{2015}]{Jones2015}
{Jones}, O.~C., {Meixner}, M., {Sargent}, B.~A., {Boyer}, M.~L., {Sewi{\l}o},
  M., \emph{et~al.} 2015, \apj, 811, 145.

\bibitem[\protect\astroncite{{Josselin} \& {Plez}}{2007}]{Josselin2007}
{Josselin}, E. \& {Plez}, B. 2007, \aap, 469, 671.

\bibitem[\protect\astroncite{{Jura} \& {Kleinmann}}{1989}]{Jura1989}
{Jura}, M. \& {Kleinmann}, S.~G. 1989, \apj, 341, 359.

\bibitem[\protect\astroncite{{Kervella} \emph{et~al.}}{2016}]{Kervella2016}
{Kervella}, P., {Lagadec}, E., {Montarg{\`e}s}, M., {Ridgway}, S.~T.,
  {Chiavassa}, A., \emph{et~al.} 2016, \aap, 585, A28.

\bibitem[\protect\astroncite{{Kervella} \emph{et~al.}}{2015}]{Kervella2015}
{Kervella}, P., {Montarg{\`e}s}, M., {Lagadec}, E., {Ridgway}, S.~T.,
  {Haubois}, X., \emph{et~al.} 2015, \aap, 578, A77.

\bibitem[\protect\astroncite{{Khouri} \emph{et~al.}}{2016}]{Khouri2016}
{Khouri}, T., {Maercker}, M., {Waters}, L.~B.~F.~M., {Vlemmings}, W.~H.~T.,
  {Kervella}, P., \emph{et~al.} 2016, \aap, 591, A70.

\bibitem[\protect\astroncite{{Klotz} \emph{et~al.}}{2012}]{klotz2012}
{Klotz}, D., {Sacuto}, S., {Kerschbaum}, F., {Paladini}, C., {Olofsson}, H.,
  \emph{et~al.} 2012, \aap, 541, A164.

\bibitem[\protect\astroncite{{Konstantinova-Antova}
  \emph{et~al.}}{2013}]{2013BlgAJ..19...14K}
{Konstantinova-Antova}, R., {Auri{\`e}re}, M., {Charbonnel}, C., {Wade}, G.,
  {Kolev}, D., \emph{et~al.} 2013, Bulgarian Astronomical Journal, 19, 14.

\bibitem[\protect\astroncite{{Lacour} \emph{et~al.}}{2009}]{lacour2009}
{Lacour}, S., {Thi{\'e}baut}, E., {Perrin}, G., {Meimon}, S., {Haubois}, X.,
  \emph{et~al.} 2009, \apj, 707, 632.

\bibitem[\protect\astroncite{{Le Bouquin} \emph{et~al.}}{2009}]{lebouquin2009}
{Le Bouquin}, J.-B., {Lacour}, S., {Renard}, S., {Thi{\'e}baut}, E., {Merand},
  A., \emph{et~al.} 2009, \aap, 496, L1.

\bibitem[\protect\astroncite{{Little-Marenin} \&
  {Little}}{1990}]{Little-Marenin1990}
{Little-Marenin}, I.~R. \& {Little}, S.~J. 1990, \aj, 99, 1173.

\bibitem[\protect\astroncite{{Maercker} \emph{et~al.}}{2012}]{Maercker2012}
{Maercker}, M., {Mohamed}, S., {Vlemmings}, W.~H.~T., {Ramstedt}, S.,
  {Groenewegen}, M.~A.~T., \emph{et~al.} 2012, \nat, 490, 232.

\bibitem[\protect\astroncite{{Marigo} \emph{et~al.}}{2013}]{marigo13}
{Marigo}, P., {Bressan}, A., {Nanni}, A., {Girardi}, L., \& {Pumo}, M.~L. 2013,
  \mnras, 434, 488.

\bibitem[\protect\astroncite{{Mart{\'{\i}}-Vidal}
  \emph{et~al.}}{2011}]{Marti-Vidal2011}
{Mart{\'{\i}}-Vidal}, I., {Marcaide}, J.~M., {Quirrenbach}, A., {Ohnaka}, K.,
  {Guirado}, J.~C., \emph{et~al.} 2011, \aap, 529, A115.

\bibitem[\protect\astroncite{{Mattsson} \emph{et~al.}}{2010}]{Mattsson2010}
{Mattsson}, L., {Wahlin}, R., \& {H{\"o}fner}, S. 2010, \aap, 509, A14.

\bibitem[\protect\astroncite{{Mayer} \emph{et~al.}}{2014}]{mayer2014}
{Mayer}, A., {Jorissen}, A., {Paladini}, C., {Kerschbaum}, F., {Pourbaix}, D.,
  \emph{et~al.} 2014, \aap, 570, A113.

\bibitem[\protect\astroncite{{Meixner} \emph{et~al.}}{2006}]{Meixner2006}
{Meixner}, M., {Gordon}, K.~D., {Indebetouw}, R., {Hora}, J.~L., {Whitney}, B.,
  \emph{et~al.} 2006, \aj, 132, 2268.

\bibitem[\protect\astroncite{{Meixner} \emph{et~al.}}{2013}]{Meixner2013}
{Meixner}, M., {Panuzzo}, P., {Roman-Duval}, J., {Engelbracht}, C., {Babler},
  B., \emph{et~al.} 2013, \aj, 146, 62.

\bibitem[\protect\astroncite{{Mohamed} \& {Podsiadlowski}}{2007}]{Mohamed2007}
{Mohamed}, S. \& {Podsiadlowski}, P. 2007, In \emph{15th European Workshop on
  White Dwarfs}, edited by R.~{Napiwotzki} \& M.~R. {Burleigh},
  \emph{Astronomical Society of the Pacific Conference Series}, vol. 372, p.
  397.

\bibitem[\protect\astroncite{{Mohamed} \& {Podsiadlowski}}{2012}]{Mohamed2012}
{Mohamed}, S. \& {Podsiadlowski}, P. 2012, Baltic Astronomy, 21, 88.

\bibitem[\protect\astroncite{{Monnier} \emph{et~al.}}{2014}]{monnier2014}
{Monnier}, J.~D., {Berger}, J.-P., {Le Bouquin}, J.-B., {Tuthill}, P.~G.,
  {Wittkowski}, M., \emph{et~al.} 2014, In \emph{Optical and Infrared
  Interferometry IV}, \emph{\procspie}, vol. 9146, p. 91461Q.

\bibitem[\protect\astroncite{{Montarg{\`e}s}
  \emph{et~al.}}{2015}]{2015EAS....71..243M}
{Montarg{\`e}s}, M., {Kervella}, P., {Perrin}, G., {Chiavassa}, A., \&
  {Auri{\`e}re}, M. 2015, In \emph{EAS Publications Series}, \emph{EAS
  Publications Series}, vol.~71, pp. 243--247.

\bibitem[\protect\astroncite{{Montarg{\`e}s}
  \emph{et~al.}}{2016}]{Montarges2016}
{Montarg{\`e}s}, M., {Kervella}, P., {Perrin}, G., {Chiavassa}, A., {Le
  Bouquin}, J.-B., \emph{et~al.} 2016, \aap, 588, A130.

\bibitem[\protect\astroncite{{Nordhaus} \& {Blackman}}{2006}]{Nordhaus2006}
{Nordhaus}, J. \& {Blackman}, E.~G. 2006, \mnras, 370, 2004.

\bibitem[\protect\astroncite{{O'Brien} \emph{et~al.}}{2006}]{OBrien2006}
{O'Brien}, T.~J., {Bode}, M.~F., {Porcas}, R.~W., {Muxlow}, T.~W.~B., {Eyres},
  S.~P.~S., \emph{et~al.} 2006, \nat, 442, 279.

\bibitem[\protect\astroncite{{O'Gorman} \emph{et~al.}}{2015}]{Gorman2015}
{O'Gorman}, E., {Vlemmings}, W., {Richards}, A.~M.~S., {Baudry}, A., {De Beck},
  E., \emph{et~al.} 2015, \aap, 573, L1.

\bibitem[\protect\astroncite{{Ohnaka}}{2014}]{Ohnaka2014}
{Ohnaka}, K. 2014, \aap, 568, A17.

\bibitem[\protect\astroncite{{Ohnaka} \emph{et~al.}}{2008}]{ohnaka2008a}
{Ohnaka}, K., {Driebe}, T., {Hofmann}, K.-H., {Weigelt}, G., \& {Wittkowski},
  M. 2008, \aap, 484, 371.

\bibitem[\protect\astroncite{{Ohnaka} \emph{et~al.}}{2016}]{Ohnaka2016}
{Ohnaka}, K., {Weigelt}, G., \& {Hofmann}, K.-H. 2016, \aap, 589, A91.

\bibitem[\protect\astroncite{{Olofsson} \emph{et~al.}}{1988}]{Olofsson1988}
{Olofsson}, H., {Eriksson}, K., \& {Gustafsson}, B. 1988, \aap, 196, L1.

\bibitem[\protect\astroncite{{Olofsson} \emph{et~al.}}{2010}]{Olofsson2010}
{Olofsson}, H., {Maercker}, M., {Eriksson}, K., {Gustafsson}, B., \&
  {Sch{\"o}ier}, F. 2010, \aap, 515, A27.

\bibitem[\protect\astroncite{{Paladini} \emph{et~al.}}{2012}]{paladini2012}
{Paladini}, C., {Sacuto}, S., {Klotz}, D., {Ohnaka}, K., {Wittkowski}, M.,
  \emph{et~al.} 2012, \aap, 544, L5.

\bibitem[\protect\astroncite{{Patat} \emph{et~al.}}{2007}]{Patat2007}
{Patat}, F., {Chandra}, P., {Chevalier}, R., {Justham}, S., {Podsiadlowski},
  P., \emph{et~al.} 2007, Science, 317, 924.

\bibitem[\protect\astroncite{{Patat} \emph{et~al.}}{2011}]{Patat2011}
{Patat}, F., {Chugai}, N.~N., {Podsiadlowski}, P., {Mason}, E., {Melo}, C.,
  \emph{et~al.} 2011, \aap, 530, A63.

\bibitem[\protect\astroncite{{Petit} \emph{et~al.}}{2013}]{2013LNP...857..231P}
{Petit}, P., {Auri{\`e}re}, M., {Konstantinova-Antova}, R., {Morgenthaler}, A.,
  {Perrin}, G., \emph{et~al.} 2013, In \emph{Lecture Notes in Physics, Berlin
  Springer Verlag}, edited by J.-P. {Rozelot} \& C.~. {Neiner}, \emph{Lecture
  Notes in Physics, Berlin Springer Verlag}, vol. 857, p. 231.

\bibitem[\protect\astroncite{{Ragland} \emph{et~al.}}{2008}]{ragland2008}
{Ragland}, S., {Le Coroller}, H., {Pluzhnik}, E., {Cotton}, W.~D., {Danchi},
  W.~C., \emph{et~al.} 2008, \apj, 679, 746-761.

\bibitem[\protect\astroncite{{Ramstedt} \emph{et~al.}}{2014}]{Ramstedt2014}
{Ramstedt}, S., {Mohamed}, S., {Vlemmings}, W.~H.~T., {Maercker}, M., {Montez},
  R., \emph{et~al.} 2014, \aap, 570, L14.

\bibitem[\protect\astroncite{{Rau} \emph{et~al.}}{2015}]{rau15}
{Rau}, G., {Paladini}, C., {Hron}, J., {Aringer}, B., {Groenewegen}, M.~A.~T.,
  \emph{et~al.} 2015, \aap, 583, A106.

\bibitem[\protect\astroncite{{Ravi} \emph{et~al.}}{2011}]{ravi2011}
{Ravi}, V., {Wishnow}, E.~H., {Townes}, C.~H., {Lockwood}, S., {Mistry}, H.,
  \emph{et~al.} 2011, \apj, 740, 24.

\bibitem[\protect\astroncite{{Richards} \emph{et~al.}}{2014}]{Richards2014}
{Richards}, A.~M.~S., {Impellizzeri}, C.~M.~V., {Humphreys}, E.~M., {Vlahakis},
  C., {Vlemmings}, W., \emph{et~al.} 2014, \aap, 572, L9.

\bibitem[\protect\astroncite{{Riebel} \emph{et~al.}}{2012}]{Riebel2012}
{Riebel}, D., {Srinivasan}, S., {Sargent}, B., \& {Meixner}, M. 2012, \apj,
  753, 71.

\bibitem[\protect\astroncite{{Sacuto} \emph{et~al.}}{2013}]{sacuto2013}
{Sacuto}, S., {Ramstedt}, S., {H{\"o}fner}, S., {Olofsson}, H., {Bladh}, S.,
  \emph{et~al.} 2013, \aap, 551, A72.

\bibitem[\protect\astroncite{{Samus} \emph{et~al.}}{2009}]{Samus2009}
{Samus}, N.~N., {Durlevich}, O.~V., \& {et al.} 2009, VizieR Online Data
  Catalog, 1, 2025.

\bibitem[\protect\astroncite{{Sargent} \emph{et~al.}}{2011}]{Sargent2011}
{Sargent}, B.~A., {Srinivasan}, S., \& {Meixner}, M. 2011, \apj, 728, 93.

\bibitem[\protect\astroncite{{Schwarzschild}}{1975}]{schwarzschild1975}
{Schwarzschild}, M. 1975, \apj, 195, 137.

\bibitem[\protect\astroncite{{Seale} \emph{et~al.}}{2014}]{Seale2014}
{Seale}, J.~P., {Meixner}, M., {Sewi{\l}o}, M., {Babler}, B., {Engelbracht},
  C.~W., \emph{et~al.} 2014, \aj, 148, 124.

\bibitem[\protect\astroncite{{Simon} \emph{et~al.}}{2009}]{Simon2009}
{Simon}, J.~D., {Gal-Yam}, A., {Gnat}, O., {Quimby}, R.~M., {Ganeshalingam},
  M., \emph{et~al.} 2009, \apj, 702, 1157.

\bibitem[\protect\astroncite{{Soker}}{2004}]{Soker2004}
{Soker}, N. 2004, In \emph{Asymmetrical Planetary Nebulae III: Winds, Structure
  and the Thunderbird}, edited by M.~{Meixner}, J.~H. {Kastner}, B.~{Balick},
  \& N.~{Soker}, \emph{Astronomical Society of the Pacific Conference Series},
  vol. 313, p. 562.

\bibitem[\protect\astroncite{{Srinivasan} \emph{et~al.}}{2016}]{Srinivasan2016}
{Srinivasan}, S., {Boyer}, M.~L., {Kemper}, F., {Meixner}, M., {Sargent},
  B.~A., \emph{et~al.} 2016, \mnras, 457, 2814.

\bibitem[\protect\astroncite{{Srinivasan} \emph{et~al.}}{2011}]{Srinivasan2011}
{Srinivasan}, S., {Sargent}, B.~A., \& {Meixner}, M. 2011, \aap, 532, A54.

\bibitem[\protect\astroncite{{Sternberg} \emph{et~al.}}{2011}]{Sternberg2011}
{Sternberg}, A., {Gal-Yam}, A., {Simon}, J.~D., {Leonard}, D.~C., {Quimby},
  R.~M., \emph{et~al.} 2011, Science, 333, 856.

\bibitem[\protect\astroncite{{Tsuji} \emph{et~al.}}{1997}]{Tsuji1997}
{Tsuji}, T., {Ohnaka}, K., {Aoki}, W., \& {Yamamura}, I. 1997, \aap, 320, L1.

\bibitem[\protect\astroncite{{van Belle} \emph{et~al.}}{2013}]{vanbelle2013}
{van Belle}, G.~T., {Paladini}, C., {Aringer}, B., {Hron}, J., \& {Ciardi}, D.
  2013, \apj, 775, 45.

\bibitem[\protect\astroncite{{Vlemmings}}{2014}]{2014IAUS..302..389V}
{Vlemmings}, W.~H.~T. 2014, In \emph{Magnetic Fields throughout Stellar
  Evolution}, edited by P.~{Petit}, M.~{Jardine}, \& H.~C. {Spruit}, \emph{IAU
  Symposium}, vol. 302, pp. 389--397.

\bibitem[\protect\astroncite{{Wallstr{\"o}m}
  \emph{et~al.}}{2015}]{Wallstrom2015}
{Wallstr{\"o}m}, S.~H.~J., {Muller}, S., {Lagadec}, E., {Black}, J.~H.,
  {Oudmaijer}, R.~D., \emph{et~al.} 2015, \aap, 574, A139.

\bibitem[\protect\astroncite{{Whitelock} \emph{et~al.}}{2009}]{Whitelock2009}
{Whitelock}, P.~A., {Menzies}, J.~W., {Feast}, M.~W., {Matsunaga}, N.,
  {Tanab{\'e}}, T., \emph{et~al.} 2009, \mnras, 394, 795.

\bibitem[\protect\astroncite{{Wittkowski} \emph{et~al.}}{2016}]{Wittkowski2016}
{Wittkowski}, M., {Chiavassa}, A., {Freytag}, B., {Scholz}, M., {H{\"o}fner},
  S., \emph{et~al.} 2016, \aap, 587, A12.

\bibitem[\protect\astroncite{{Wittkowski} \& {Paladini}}{2014}]{wittkowski2014}
{Wittkowski}, M. \& {Paladini}, C. 2014, In \emph{EAS Publications Series},
  \emph{EAS Publications Series}, vol.~69, pp. 179--195.

\bibitem[\protect\astroncite{{Woitke}}{2006}]{Woitke2006}
{Woitke}, P. 2006, \aap, 460, L9.

\bibitem[\protect\astroncite{{Wong} \emph{et~al.}}{2016}]{Wong2016}
{Wong}, K.~T., {Kami{\'n}ski}, T., {Menten}, K.~M., \& {Wyrowski}, F. 2016,
  \aap, 590, A127.

\bibitem[\protect\astroncite{{Wood}}{1990}]{Wood1990}
{Wood}, P.~R. 1990, In \emph{Confrontation Between Stellar Pulsation and
  Evolution}, edited by C.~{Cacciari} \& G.~{Clementini}, \emph{Astronomical
  Society of the Pacific Conference Series}, vol.~11, pp. 355--363.

\end{thebibliography}

\end{document}